\documentclass[usenatbib]{mn2e}

\usepackage{natbib}
\usepackage[utf8x]{inputenc}
\usepackage{amsmath}
\usepackage{bm}
\usepackage{booktabs}
\usepackage{graphicx}

\newcommand       \aap          {A\&A }
\newcommand       \ssr           {Space Sci. Rev.}

\newcommand       \jgr       {J. Geophys. Res.}

\newcommand       \apj        {ApJ}
\newcommand       \apjl        {ApJL}
\newcommand       \mnras   {MNRAS}
\newcommand       \memsai {MemSAIt}

\newcommand       \physrep  {Phys. Rep.}
\newcommand       \prd         {Phys. Rev. D}
\newcommand       \pre         {Phys. Rev. E}

\newcommand{\be}{\begin{equation}}
\newcommand{\ee}{\end{equation}}
\newcommand{\bdm}{\begin{displaymath}}
\newcommand{\edm}{\end{displaymath}}
\newcommand{\bea}{\begin{eqnarray}}
\newcommand{\eea}{\end{eqnarray}}
\newcommand{\ba}{\begin{align}}
\newcommand{\ea}{\end{align}}

\title[Fast reconnection in relativistic plasmas]
{Fast reconnection in relativistic plasmas: \\ the magnetohydrodynamics tearing instability revisited}

\author[L. Del Zanna et al.]
{L. Del Zanna$^{1,2,3}$\thanks{luca.delzanna@unifi.it}, E. Papini$^4$, S. Landi$^{1,2}$, M. Bugli$^{5,6}$, N. Bucciantini$^{2,1,3}$ \\
$^1$ Dipartimento di Fisica e Astronomia, Università di Firenze, Via G. Sansone 1, 50019 Sesto F.no (Firenze), Italy \\
$^2$ INAF -- Osservatorio Astrofisico di Arcetri, L.go E. Fermi 5, 50125 Firenze, Italy\\
$^3$ INFN -- Sezione di Firenze, Via G. Sansone 1, 50019 Sesto F.no (Firenze), Italy\\
$^4$ Max-Planck-Institut f\"ur Sonnensystemforschung, J.~von Liebig Weg 3, 37077 G\"ottingen, Germany\\
$^5$ Max-Planck-Institut f\"ur Astrophysik, Karl-Schwartzschild Strasse 1, 85741 Garching, Germany\\
$^6$ Technische Universit\"at M\"unchen,   Arcisstrasse 21, 80333 M\"unchen, Germany
}
\date{}

\begin{document}

\maketitle
\begin{abstract}
Fast reconnection operating in magnetically dominated plasmas is often invoked in models for magnetar giant flares, for magnetic dissipation in pulsar winds, or to explain the gamma-ray flares observed in the Crab nebula, hence its investigation is of paramount importance in high-energy astrophysics. Here we study, by means of two dimensional numerical simulations, the linear phase and the subsequent nonlinear evolution of the tearing instability within the framework of relativistic resistive magnetohydrodynamics, as appropriate in situations where the Alfv\'en velocity approaches the speed of light. It is found that the linear phase of the instability closely matches the analysis in classical MHD, where the growth rate scales with the Lundquist number $S$ as $S^{-1/2}$, with the only exception of an enhanced inertial term due to the thermal and magnetic energy contributions. In addition, when thin current sheets of inverse aspect ratio scaling as $S^{-1/3}$ are considered, the so-called \emph{ideal} tearing regime is retrieved, with modes growing independently on $S$ and extremely fast, on only a few light crossing times of the sheet length. The overall growth of fluctuations is seen to solely depend on the value of the background Alfv\'en velocity. In the fully nonlinear stage we observe an inverse cascade towards the fundamental mode, with Petschek-type supersonic jets propagating at the external Alfv\'en speed from the X-point, and a fast reconnection rate at the predicted value $\mathcal{R}\sim (\ln S)^{-1}$. 
\end{abstract}

\section{Introduction}

Magnetic reconnection is the primary process through which the energy contained in magnetically dominated plasmas can be transferred rapidly into heat, high-speed plasma flows and particle acceleration, for example during the flare activity observed in magnetic structures in the solar corona, in geomagnetic substorms, or in laboratory plasmas. Reconnection is a change of topology of magnetic fields, typically occurring in elongated sheets where Ohmic or other dissipative effects, usually negligible, are enhanced due to the local strong intensity of the electric current. Within the fluid approximation of magnetohydrodynamics (MHD) and for non-relativistic velocities, the time evolution of the magnetic field is governed by the resistive induction equation: we have competition between the ideal timescale $\tau_A=L/c_A$ and the diffusion timescale $\tau_D=L^2/\eta$, where $L$ is a characteristic scale of variation of the equilibrium magnetic field, $c_A$ the Alfv\'en speed and $\eta$ the magnetic diffusivity. Classical models for reconnection, like the nonlinear steady-state model by Sweet-Parker \citep{Sweet:1958,Parker:1957} and the linear analysis of the tearing instability \citep{Furth:1963}, predict reconnection timescales behaving as $\tau=\sqrt{\tau_A\tau_D}$, that increases with the square root of the Lunquist number $S=\tau_D/\tau_A=Lc_A/\eta$. It is well-known that in astrophysical or laboratory highly conducting plasmas we typically estimate $S\gg 1$, for example $S=10^{12}$ in the solar corona, therefore classical models fail to predict the observed bursty phenomena such as solar flares or tokamak disruptions, which occur instead on a non-negligible fraction (say not less than $\sim10\%$) of the ideal (Alfv\'enic) timescales. 

To overcome these difficulties, in the last decades the plasma physics community has focussed  on non-MHD effects believed to occur at the small scales where reconnection takes place, like separation of scales between protons and electrons or other kinetic effects \citep[e.g.][]{Yamada:2010}. However, it has been recently realized that even within single-fluid MHD, provided the Lundquist number is high enough and the current sheet aspect ratio $L/a$ is also large (from now on $L$ and $a$ indicate the sheet length and width, respectively), the tearing mode can become very fast. Moreover, in this case explosive secondary reconnection events and production of plasmoid chains on smaller and smaller scales are observed (sometimes named as \emph{super-tearing} or \emph{plasmoid instability}), leading to a reconnection rate of approximately $\mathcal{R} \sim 10^{-2}$, basically independent on $S$ \citep{Loureiro:2007,Lapenta:2008,Samtaney:2009,Bhattacharjee:2009,Cassak:2009a,Huang:2010a,Uzdensky:2010,Cassak:2012,Loureiro:2013,Loureiro:2016}. In particular, the Sweet-Parker (SP) current sheet, of inverse aspect ratio $a/L\sim S^{-1/2}$ as derived in the stationary 2-D model for driven reconnection, was shown to be tearing unstable with growth rates $\gamma \tau_A\sim S^{1/4}$, occurring on very fast timescales provided $S>S_c\simeq 10^4$. This is clearly a paradoxical result,  since the ideal limit, corresponding to $S\to\infty$, would lead to infinitely fast instabilities, while it is known that in ideal MHD the flux-freezing theorem holds and reconnection is impossible. 

The answer to this paradox was provided by \citet{Pucci:2014}, who studied the tearing mode instability in current sheets with generic inverse aspect ratios $a/L\sim S^{-\alpha}$, finding that there exists a threshold, for $\alpha = 1/3$, separating slow and fast reconnection. At this critical value, the growth rate of the instability becomes asymptotically independent of $S$ (provided $S \sim 10^6$ or greater) and $\gamma\tau_A \simeq 0.6$, so that we can speak of an \emph{ideal} tearing mode. We then expect that during the current sheet evolution, as long as $\alpha < 1/3$ the instability remains too slow to affect the overall dynamics, but as soon as the aspect ratio $L/a\sim S^{1/3}$ is reached in the thinning process, the ideal tearing mode sets in, leading to fast magnetic dissipation and finally to the flare event. Notice that the SP current sheet, which is thinner, cannot exist in nature within the proposed scenario. The ideal tearing instability thus solves, at the same time and within the macroscopic resistive MHD regime, both the paradox of the \emph{super-tearing} instability in SP current sheets and the quest for fast magnetic spontaneous reconnection, so that observations of the explosive events occurring in space, astrophysical, and laboratory plasmas can be matched.

This idea has been first proved numerically in \cite{Landi:2015}, who investigated the stability of a current sheet with $a/L = S^{-1/3}$ via 2-D, compressible MHD simulations. In their work the characteristic eigenfunctions of the tearing instability are retrieved with high precision, the expected dispersion relation is found for a variety of values of $S$ and of the plasma $\beta$, in both pressure and force-free equilibria,  and the asymptotic fast rate of  $\gamma\tau_A \simeq 0.6$ is measured for the expected fastest growing modes. In addition, in the fully nonlinear evolutionary stage, the authors observe the merging of nearby plasmoids and secondary reconnection events, occurring when the condition $a/L\sim S^{-1/3}$ is reached locally on the newly formed and smaller reconnecting site, leading to the explosive expulsion of smaller-scale plasmoids from the X-point (additional details in \cite{Del-Zanna:2016}). A dynamical scenario of converging magnetic structures producing elongated, collapsing current sheets has been investigated numerically by \cite{Tenerani:2015a}, who confirmed that the ideal tearing sets in when the critical threshold is reached, secondary plasmoid instabilities soon disrupt the current sheet, and the SP configuration is never reached.

What discussed so far applies to classical MHD. However, astrophysical plasmas are often magnetically dominated, in the sense that fields are so strong that the Alfv\'en speed may approach the speed of light. It is well known that Soft Gamma Repeaters and Anomalous X-ray Pulsars are most likely activated by huge flares of magnetars, neutron stars with fields in their magnetospheres up to a thousand times stronger than in standard radio pulsars, say $B\sim 10^{15}$~G \citep{Usov:1994,Thompson:1995,Lyutikov:2006a,Elenbaas:2016}. Fast-spinning newly born magnetars are becoming a favourite model for the inner engine of both long and short Gamma-Ray Bursts \citep{Komissarov:2007a,Bucciantini:2009,Metzger:2011,Bucciantini:2012}. Magnetically dominated plasmas are also expected around black holes, probably originated in the coronae of their accretion disks, and give rise to jet launching both in AGN/microquasar systems \citep{Romanova:1992,Barkov:2012,Sironi:2015} and in Gamma-Ray Bursts \citep{Drenkhahn:2002,Barkov:2008,Kumar:2015}. 

In addition, relativistic reconnection could be important in pair plasmas \citep[see][and references therein]{Kagan:2015}, and in particular it should be an ingredient in the modeling of Pulsar Wind Nebulae in order to explain dissipation of magnetic energy in the so-called \emph{striped wind} region \citep{Coroniti:1990,Kirk:2003,Petri:2007,Amano:2013,Takamoto:2015}, and the sudden gamma-ray flares observed in the Crab nebula, probably originated in the vicinities of the wind termination shock \citep{Tavani:2013,Baty:2013,Cerutti:2014, Lyutikov:2016a}. Promising results are starting to come from two-fluid relativistic MHD simulations \citep{Barkov:2014,Barkov:2016}, as a viable alternative to fully kinetic particle-in-cell (PIC) methods, which are more computationally demanding.

When the magnetic field energy is dominant over other forms of energy, the instability growth and the reconnection rates were initially believed to be greatly enhanced \citep{Blackman:1994,Lyutikov:2003b}. However, it was later suggested that, in spite of the presence of a non-vanishing charge density and the displacement current in the relativistic equations, the classical models for 2-D reconnection apply substantially unchanged even in this regime. \cite{Lyubarsky:2005} analyzed analytically the SP and Petschek \citep{Petschek:1964} stationary models for a relativistic plasma, while \cite{Komissarov:2007b} studied both analytically and numerically the tearing mode in the so-called force-free degenerate (or \emph{magnetodynamics}) regime, in which fluid velocities and energy density are negligible with respect to the drift speed and the magnetic energy density, respectively. Surprisingly, these studies basically confirmed the classic results, once the Alfv\'en velocity is replaced by the speed of light. 

Additional numerical simulations of tearing unstable SP current sheets within relativistic magnetodynamics or MHD have basically confirmed this scenario \citep[e.g.][]{Watanabe:2006,Zenitani:2010,Takahashi:2013}. In these works secondary instabilities and formation of plasmoid chains are observed, as in \citet{Samtaney:2009} and subsequent works in classical MHD. In some cases Petschek-like configurations are also found during the nonlinear stage, possibly due to the assumption of a non-uniform enhanced resistivity. A comprehensive study of nonlinear fast reconnection in relativistic MHD was performed by \cite{Takamoto:2013}, who investigated high Lundquist number and high magnetization regimes up to $c_A=0.98c$. In that work, explosive formation and evolution of plasmoid chains was observed, an accurate investigation of the motion of X and O-points was performed together with a statistical analysis of the number and size of plasmoids, in the spirit of the stochastic model by \cite{Uzdensky:2010}. The reconnection rate was found to saturate for $S>S_c\simeq 10^4$, and formation of a \emph{monster} plasmoid continuously fed by smaller ones was retrieved. 2-D and 3-D simulations were performed at even higher Lundquist numbers by \citet{Zanotti:2011}, who employed Galerkin methods on unstructured meshes. They observed outflows from the X-point with higher speed at increasing magnetization, and claimed that the critical Lundquist number to observe the explosive plasmoid instability raises to $S_c\sim 10^8$ in the relativistic case.

What still deserves investigation is certainly the extension to the relativistic MHD regime of the ideal tearing model. Since in this scenario SP current sheets are not expected to form during dynamical evolution (the critical aspect ratio is reached before the SP one, as previously discussed), the whole picture of the tearing instability needs to be revisited also for magnetically dominated plasmas. 

In this work we shall present for the first time relativistic MHD simulations of the tearing instability in current sheets with thickness $a=S^{-1/3}L$, extending the analysis by \cite{Landi:2015} to strongly magnetized plasmas with $c_A\simeq c$. A linear analysis of the instability will be first performed, retrieving the appropriate scalings with $S$ and the full dispersion relation, both analytically and numerically. Then, the nonlinear phase of the instability will be followed up to a quasi-steady Petschek-type configuration. We shall  demonstrate that the full evolution just requires $\sim 10-20$ light crossing times of the sheet length, thus proving that extremely fast reconnection is indeed possible in strongly magnetized plasmas, as required to explain many observed explosive events in high-energy astrophysics, as the giant flares of magnetars or those from the Crab nebula.

The paper is structured as follows: in section~\ref{sect:model} we review the relativistic resistive MHD equations and provide details on the numerical methods, in section~\ref{sect:standard} we propose a semi-relativistic treatment for the tearing mode in magnetically dominated plasmas, extended to sheets with $a/L\sim S^{-1/3}$ in section~\ref{sect:ideal}. Numerical simulations are presented in section~\ref{sect:results_linear} (linear, single-mode runs) and \ref{sect:results_nonlinear} (fully nonlinear runs). Conclusions and discussion of a couple of natural astrophysical applications of our model are reported in section~\ref{sect:concl}.

\section{Equations and numerical scheme}
\label{sect:model}

In the present section we describe the equations for resistive relativistic MHD as implemented in the ECHO code \citep{Del-Zanna:2007} and employed here for the numerical simulations of the tearing instability.  Assuming a flat Minkowskian metric, the evolution equations for a system composed by a (single) fluid and electromagnetic fields are (we let $c\to 1$, $4\pi\to 1$):
\bea
& & \!\!\!  \partial_t D + \bm{\nabla} \cdot ( \rho\Gamma \bm{v} ) = 0, \\
& &  \!\!\!  \partial_t \bm{S} + \bm{\nabla} \cdot (w \Gamma^2 \bm{v}\bm{v} 
- \bm{E}\bm{E}-\bm{B}\bm{B} + p_\mathrm{tot}\bm{I} ) = 0, \\
& &   \!\!\!  \partial_t {\mathcal E} + \bm{\nabla} \cdot ( w \Gamma^2 \bm{v} + \bm{E} \times \bm{B} ) = 0, \\
& &  \!\!\!  \partial_t \bm{B} + \bm{\nabla}\times\bm{E} = 0, \label{eq:magnetic} \\
& &  \!\!\!  \partial_t \bm{E} - \bm{\nabla}\times\bm{B} = - \bm{J}, \label{eq:electric}
\eea
where $\rho$ is the rest mass density, $p$ the thermal pressure, $w=e+p=\rho+4p$ the enthalpy density for an ideal gas with adiabatic index of $4/3$, $e$ being the energy density (for an ultrarelativistic fluid with $p=e/3\gg\rho$ we may let $w=4p$). All these quantities refer to the \emph{comoving frame} of the plasma. In the \emph{laboratory frame} we define $\bm{v}$ and $\Gamma=1/\sqrt{1-v^2}$ as the fluid bulk velocity and the corresponding Lorentz factor, $\bm{E}$ and $\bm{B}$ are the electric and magnetic fields, $u_\mathrm{em}=\tfrac{1}{2}(E^2+B^2)$ their energy density, and $p_\mathrm{tot}=p+u_\mathrm{em}$. We have also defined the following conservative variables, again measured in the laboratory frame: the rest mass density $D=\rho\Gamma$, the total momentum density  $\bm{S}=w \Gamma^2 \bm{v} + \bm{E} \times \bm{B}$, and the total energy density ${\mathcal E} =w \Gamma^2 - p + u_\mathrm{em}$, whereas $\bm{J}$ is the current density, which appears as a source term. These evolution equations for 11 physical variables must be supplemented by the remaining Maxwell equations, that is the divergence-less condition for the magnetic field $\bm{\nabla}\cdot\bm{B}=0$ and the Gauss law $\bm{\nabla}\cdot\bm{E}=\rho_\mathrm{e}$, where $\rho_\mathrm{e}$ is the charge density, and by a closure for the propagation of currents in the medium, namely some form of Ohm's law.

The fully covariant formulation for the Ohm law in a resistive plasma, including the effects of mean-field dynamo action, was proposed in \citet{Bucciantini:2013}. In terms of the Minkowskian quantities defined above and considering isotropic dissipation alone, it translates into
\be
\bm{J} = \rho_\mathrm{e}\bm{v} + \eta^{-1}\, \Gamma[\bm{E}+\bm{v}\times\bm{B}-(\bm{E}\cdot\bm{v})\bm{v}],
\label{eq:ohm}
\ee
as also used in previous numerical works \citep{Komissarov:2007,Palenzuela:2009}. Here $\eta$ is the coefficient of magnetic diffusivity (or simply the resistivity), that is considered as a scalar, constant quantity in the present work. For small velocities, that is in the classic MHD case, the evolution equation~(\ref{eq:electric}) for the electric field reduces to Ampere's law $\bm{J}=\bm{\nabla}\times\bm{B}$, the above equation becomes $\bm{E} = - \bm{v}\times\bm{B} + \eta\bm{J}$, so we retrieve the fact that both $\bm{E}$ and $\bm{J}$ are derived quantities in the non-relativistic limit, and the Maxwell equation~(\ref{eq:magnetic}) reduces to the familiar induction equation for classical MHD
\be
\partial_t \bm{B} = \bm{\nabla}\times (\bm{v} \times \bm{B} ) + \eta \bm{\nabla}^2\bm{B}.
\label{eq:induction}
\ee

In typical astrophysical situations the resistivity coefficient is very small and thus equation~(\ref{eq:electric}), combined with equation~(\ref{eq:ohm}), is a \emph{stiff} equation: terms $\propto\eta^{-1}$ can evolve on timescales much faster than the ideal ones. We have then to solve a system of hyperbolic partial differential equations combined with stiff relaxation equations, that can be rewritten in the form
\be
\partial_t \bm{U} = \bm{Q} ( \bm{U} ) + \bm{R} ( \bm{U} ),
\ee
where $\bm{U}=\{D,\bm{S},{\mathcal E} ,\bm{B},\bm{E} \}$ is the set of conserved variables, $\bm{Q}$ are the non-stiff terms (divergence of fluxes, curl of EM fields, the charge density source term), whereas $\bm{R}$ are the stiff terms $\propto \eta^{-1}$ requiring some form of implicit treatment in order to preserve numerical stability. For this reason, time integration is here achieved by using the so-called \emph{IMplicit-EXplicit} (IMEX) Runge-Kutta methods developed by \cite{Pareschi:2005} and first employed in resistive relativistic MHD by \cite{Palenzuela:2009}. Further details of the implementation in the ECHO code can be found in \cite{Del-Zanna:2014} and \cite{Bugli:2014}. Given the vector $\bm{U}^n$ at time $t^n$, IMEX-RK schemes are characterized by $s$ sub-cycles
\be
\bm{U}^i = \bm{U}^n + \Delta t \sum_{j=1}^{i-1} \tilde{a}_{ij}\bm{Q}(\bm{U}^j)
+  \Delta t \sum_{j=1}^{i} a_{ij}\bm{R}(\bm{U}^j),
\ee
for $i=1,\ldots, s$, and by a final, explicit step yielding $\bm{U}^{n+1}$ at time $t^{n+1} = t^n + \Delta t$
\be
\bm{U}^{n+1} = \bm{U}^n + \Delta t \sum_{i=1}^{s} \tilde{b}_{i}\bm{Q}(\bm{U}^i)
+  \Delta t \sum_{i=1}^{s} b_{i}\bm{R}(\bm{U}^i).
\ee
Here $A=\{a_{ij}\}$ and $\tilde{A}=\{\tilde{a}_{ij}\}$ are square matrices of $s\times s$ coefficients ($a_{ij}=0$ for $j>1$ and $\tilde{a}_{ij}=0$ for $j\ge 1$), whereas $B=\{b_i\}$ and $\tilde{B}=\{\tilde{b}_i\}$ are vectors with $s$ components. Notice that the last sub-step $j=i$ in the first equation is implicit for the electric field $\bm{E}$. Indicating with $ \bm{E}_*$ the (known) sum up to index $i-1$ and letting $\tilde{\eta} = \eta / ( \Delta t \, a_{ii})$, we then solve
\be
\bm{E} = \bm{E}_* - \tilde{ \eta}^{-1}\, \Gamma[\bm{E}+\bm{v}\times\bm{B}-(\bm{E}\cdot\bm{v})\bm{v}].
\ee
The above equation can be inverted to provide $\bm{E}$, that is
\be
(1+\tilde{\eta}\,\Gamma^{-1}) \bm{E} = - \bm{v}\times\bm{B} + \tilde{\eta}\,\left[ \Gamma^{-1} \bm{E}_* + \frac{(\bm{E}_* \cdot\bm{v})\bm{v}}{ \tilde{\eta} + \Gamma^{-1} } \right] ,
\ee
which nicely reduces to the ideal case when $\eta=0\Rightarrow\tilde{\eta}=0$. This inversion is nested inside the iterative Newton-Raphson scheme to retrieve primitive variables from the set of conservative ones, as described in \cite{Bucciantini:2013}. In the present paper we use the third order, four stages ($s=4$) strong stability preserving SPP3(4,3,3) IMEX-RK scheme, as described in the cited papers. Spatial integration is achieved by using the MP5 limited reconstruction (up to fifth order of accuracy for smooth solutions) combined with the HLL two-wave Riemann solver, and the divergence-less constraint for $\bm{B}$ is enforced via the UCT method. For the explanation of these numerical techniques and further details see \cite{Del-Zanna:2007}. 

\section{The tearing instability in relativistic MHD: linear analysis}
\label{sect:standard}

The tearing mode is studied here by analyzing the growth of the perturbations of an initial force-free current sheet in which the magnetic field $\bm{B}_0$ varies in the $x$ direction over a characteristic scale $a$ (e.g. $\sim\tanh(x/a)$, with $a$ being the half width) and is elongated along $y$.  A $z$ component of the field, concentrated near the sheet axis, is also needed to ensure a rotation in the $y-z$ plane across the sheet preserving an overall constant magnitude $B_0$. This is the case $\zeta=1$ of the more general equilibrium used by \cite{Landi:2015}. In the present work we then assume
\be
\bm{B}_0 = B_0 [ \tanh (x/a) \hat{\mathbf{y}} + \mathrm{sech} (x/a) \hat{\mathbf{z}} ].
\label{eq:b0}
\ee
Since $|\bm{B}_0|=B_0=\mathrm{const}$, also the equilibrium density and pressure, $\rho_0$ and $p_0$ respectively, are constant for the force-free equilibrium assumed here. We choose to employ these quantities in the definition of two non-dimensional parameters, relevant in relativistic MHD, namely the plasma magnetization and the plasma beta
\be
\sigma_0 = B_0^2/\rho_0, \quad \beta_0=2p_0/B_0^2,
\ee
so the knowledge of $B_0$, $a$, and the above two parameters uniquely determines the initial force-free, static configuration. This is an exact equilibrium in the ideal MHD limit $\eta\to 0$, but not in the resistive case.

Consider now a quasi-equilibrium evolution phase where small, non-relativistic velocities ($v=|\bm{v}|\ll 1$, $\Gamma\simeq 1$) are allowed to perturb the initial state. The electric field is much weaker than the magnetic field (in magnitude $E \sim v B$), therefore in the energy-momentum tensor the $v^2$ and $E^2$ terms vanish, though the pressure and magnetic field contributions to the plasma inertia must be retained. Moreover, from equation~(\ref{eq:ohm}) we retrieve the already mentioned classical limit of equation~(\ref{eq:induction}) for the time evolution of $\bm{B}$. Thus, at first-order level, the induction and momentum equations can be written as
\begin{align}
& \partial_t \bm{B}_1 = \bm{\nabla}\times (\bm{v}_1 \times \bm{B}_0 ) + \eta \bm{\nabla}^2\bm{B}_1, \label{eq:linear1} \\
&  \partial_t (w_0 \bm{v}_1 + \bm{E}_1\times\bm{B}_0)  =  \nonumber \\ 
& - \bm{\nabla}( p_1 +\bm{B}_0\cdot\bm{B}_1) + (\bm{B}_0\cdot\bm{\nabla}) \bm{B}_1 + (\bm{B}_1\cdot\bm{\nabla}) \bm{B}_0,
\end{align}
where the $0$ subscripts denote equilibrium quantities, and the $1$ subscripts the perturbed ones.

If we now make the simplifying assumption that the linearized Poynting flux has as a leading contribution $\bm{E}_1\times\bm{B}_0 \simeq B_0^2\bm{v}_1$, neglecting the term $-(\bm{v}_1\cdot\bm{B}_0)\bm{B}_0$ (we have verified that during the evolution in our numerical simulations it is always much smaller than the one we retain, at least of a factor 10), and that proportional to $\eta$, we may write 
\be
\partial_t (w_0 \bm{v}_1 + \bm{E}_1\times\bm{B}_0) \simeq (w_0+B_0^2)  \partial_t \bm{v}_1,
\ee
and the equations governing the tearing instability take the same form as in the classical MHD, provided we employ $w_0+B_0^2$ instead of $\rho_0$ in the normalizations, as we will now show.

We first write the Alfv\'en speed, calculated on the equilibrium quantities, in the appropriate relativistic form
\begin{align}
& c_A =  \frac{B_0}{\sqrt{w_0 + B_0^2}} =  \frac{B_0}{\sqrt{\rho_0 + 4p_0 + B_0^2}} \nonumber \\ 
& = (1/\sigma_0 + 2\beta_0 + 1)^{-1/2},
\label{eq:ca}
\end{align}
that verifies  $c_A\leq 1$. Moreover, $c_A\to 1$ in the high magnetization limit $\sigma_0\gg 1$ and for low-beta plasmas (for an ultrarelativistic fluid with $p=e/3\gg\rho$ the $1/\sigma_0$ term vanishes). Notice the enhanced inertia, since in a relativistic plasma all energetic contributions (rest mass, thermal, and magnetic) must contribute to it.

As in the classical MHD treatment, we take the curl of the momentum equation, to remove the contribution of the thermal and magnetic pressure terms, and assume that the flow is incompressible in the linear phase of the instability. Next we consider a linear perturbation of the form
\be
{B_1}_x = b(x) \exp (\gamma t + iky), \quad {v_1}_x = v(x) \exp (\gamma t + iky), 
\ee
where $\gamma$ and $k$ are real quantities defining the growth rate and the wave number of the instability. The $y$ components are easily retrieved from the solenoidal conditions $\bm{\nabla}\cdot\bm{B}_1=0$ and $\bm{\nabla}\cdot\bm{v}_1=0$, while we further assume ${B_1}_z = {v_1}_z =0$.

It is convenient to normalize all quantities against the length-scale $a$ of the force-free equilibrium, the corresponding Alfv\'enic crossing time $\bar{\tau}_A=a/c_A$, and the equilibrium magnetic field strength, such that
\begin{align}
& \bar{x}=\frac{x}{a} , \quad \bar{y}=\frac{y}{a}, \quad \bar{k} = ka, \quad \bar{t}=\frac{t}{\bar{\tau}_A}, \quad \bar{\gamma} = \gamma\bar{\tau}_A, \nonumber \\
&  \bar{v}=\frac{v}{c_A}, \quad \bar{B}=\frac{B}{B_0}, \quad \bar{b}=\frac{b}{B_0}. 
\end{align}
With the above choices and the relativistic definition of $c_A$ in (\ref{eq:ca}), the final set of normalized equations to be solved is
\begin{align}
& \bar{\gamma}\, \bar{b} = i\bar{k}\bar{B}\bar{v} + \bar{S}^{-1}(\bar{b}^{\prime\prime} - \bar{k}^2 \bar{b}),  \label{eq:classic_tearing1} \\
& \bar{\gamma} (\bar{v}^{\prime\prime} - \bar{k}^2 \bar{v}) = i\bar{k} [ \bar{B} ( \bar{b}^{\prime\prime} - \bar{k}^2 \bar{b}) - \bar{B}^{\prime\prime} \bar{b}], 
\label{eq:classic_tearing2}
\end{align}
where primes indicate differentiation with respect to $\bar{x}$ and where we have introduced the Lundquist number relative to the length-scale $a$ as
\be
\bar{S} = \frac{ac_A}{\eta}.
\ee
As anticipated, equations~(\ref{eq:classic_tearing1}-\ref{eq:classic_tearing2}) are the same as in classical MHD, with the exception of the different inertial term in $c_A$. We then expect to find the usual dispersion relation for the tearing mode, with maximum growth rate for the fastest growing mode and corresponding maximum wavenumber  \citep{Furth:1963,Priest:2000}
\begin{align}
& \bar{\gamma}_\mathrm{max} \equiv \gamma_\mathrm{max} \bar{\tau}_A \simeq 0.6\,\bar{S}^{-1/2}, \label{eq:rates1} \\
& \bar{k}_\mathrm{max} \equiv k_\mathrm{max} a \simeq 1.4 \,\bar{S}^{-1/4}, \label{eq:rates1b}
\end{align}
where we can appreciate the standard dependences on $\bar{S}$. Notice that a similar analysis was performed in the case of force-free degenerate electrodynamics by \cite{Komissarov:2007b}, and even in this limiting case where electromagnetic forces dominate over the hydrodynamical ones, the results are the same as in classical MHD, provided the bulk fluid velocity is replaced by the drift velocity $\bm{E}\times\bm{B}/B^2$ and $c_A\to c\equiv 1$ (we notice that there is a factor 2 missing in the first term of their momentum equation~49 and in the other expressions derived from it).

Differences in the relativistic case are only apparent if we measure the instability growth rates in terms of the light crossing time $\bar{\tau}_c=a/c\equiv a$, as invariably happens when analyzing the results in a relativistic context. In this case, preserving the definition of $\bar{S}$ given above, the maximum growth rate would be measured as
\be
\gamma_\mathrm{max} \bar{\tau}_c \simeq 0.6\,c_A\,\bar{S}^{-1/2},
\label{eq:rates2}
\ee
which appears as a reduced value only because of the different normalization, here with respect to a shorter timescale $\bar{\tau}_c < \bar{\tau}_A$ (as $c_A<1$). Finally, notice that the above estimate is valid for assigned values of $\bar{S}$ and $c_A$. Thus, in simulations, the resistivity must be properly redefined as $\eta=ac_A/\bar{S}$ in order to match the desired value of $\bar{S}$.

\section{The relativistic ideal tearing instability in thin current sheets}
\label{sect:ideal}

What discussed so far is the relativistic extension of the linear phase of the tearing mode instability, applied to a force-free magnetic field equilibrium describing a current sheet of (half) width $a$. The expected growth rate $\bar{\gamma}_\mathrm{max}\sim \bar{S}^{-1/2}$ gets smaller with higher $\bar{S}$, as in classical MHD, but it is clear that this result must be interpreted carefully.  In the following we choose to measure both the time and the Lundquist number in terms of a \emph{macroscopic} lengthscale $L$, namely the length of a thin current sheet whose aspect ratio is then $L/a\gg 1$. We obtain
\be
\tau_A = \frac{L}{c_A}, \quad S = \frac{Lc_A}{\eta},
\ee
so that
\be
\bar{\tau}_A = \frac{a}{L}\tau_A \ll \tau_A, \quad \bar{S} = \frac{a}{L}S \ll S,
\label{eq:bar}
\ee
thus the growth rates are expected to increase, when normalized with respect to these macroscopic quantities. The result for thin current sheets is
\be
\gamma_\mathrm{max}\tau_A \simeq 0.6\, S^{-1/2} (a/L)^{-3/2},
\ee
which in SP sheets with $a/L\sim S^{-1/2}$ is known to lead to the paradoxical result of a rate growing with a positive power of $S$ (we recall that the ideal plasma limit should be recovered for $S\to\infty$) and to \emph{super-tearing} events  \citep[e.g.][]{Loureiro:2007,Samtaney:2009,Bhattacharjee:2009}.

Following \cite{Pucci:2014}, it is instructive to parametrize in terms of $S$ the inverse aspect ratio as $a/L\sim S^{-\alpha}$. We then find that there is a threshold critical value $\alpha=1/3$ at which the tearing growth time becomes ideal. This is because for
\be
a = L \,S^{-1/3},
\ee
the dependence on $S$ completely disappears and the relations (\ref{eq:rates1}-\ref{eq:rates1b}), with the normalizations in  (\ref{eq:bar}), lead to
\be
\gamma_\mathrm{max}\tau_A \simeq 0.6, \quad k_\mathrm{max}L \simeq 1.4 \, S^{1/6}.
\label{eq:rates3}
\ee
The instability now occurs on the (ideal) Alfv\'enic time $\tau_A$ as measured on the macroscopic scales, as it should be to reproduce astrophysical observations of reconnection events. Therefore, in a dynamical process with ever decreasing $a/L$ values, the Sweet-Parker configuration with $\alpha=1/2>1/3$ is never reached, given that a thin sheet with $a/L\sim S^{-1/3}$ is already unstable on ideal timescales. 


\begin{figure*}
\centerline{
\hspace{ -8mm} \includegraphics[width=72mm]{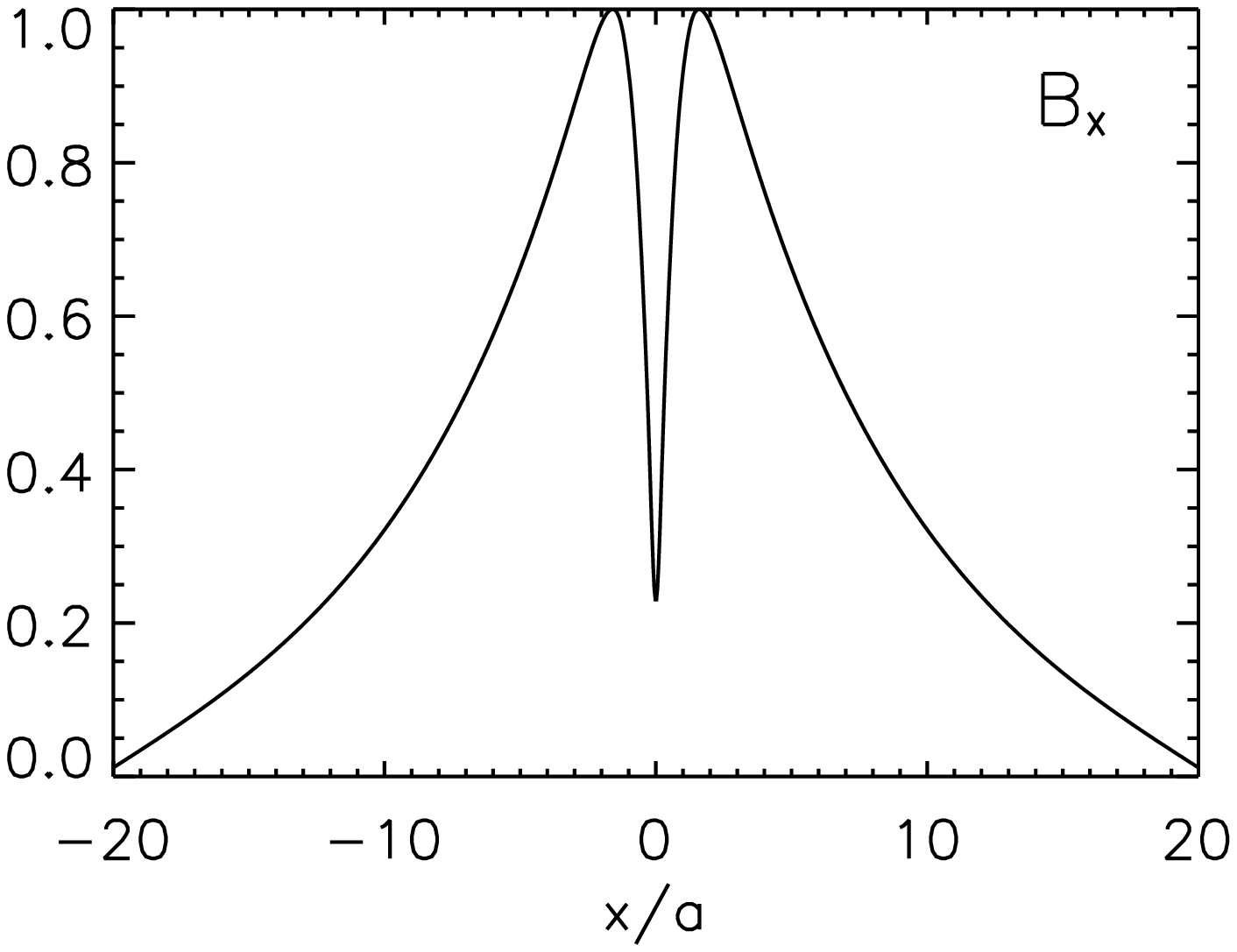} 
\hspace{-14mm} \includegraphics[width=72mm]{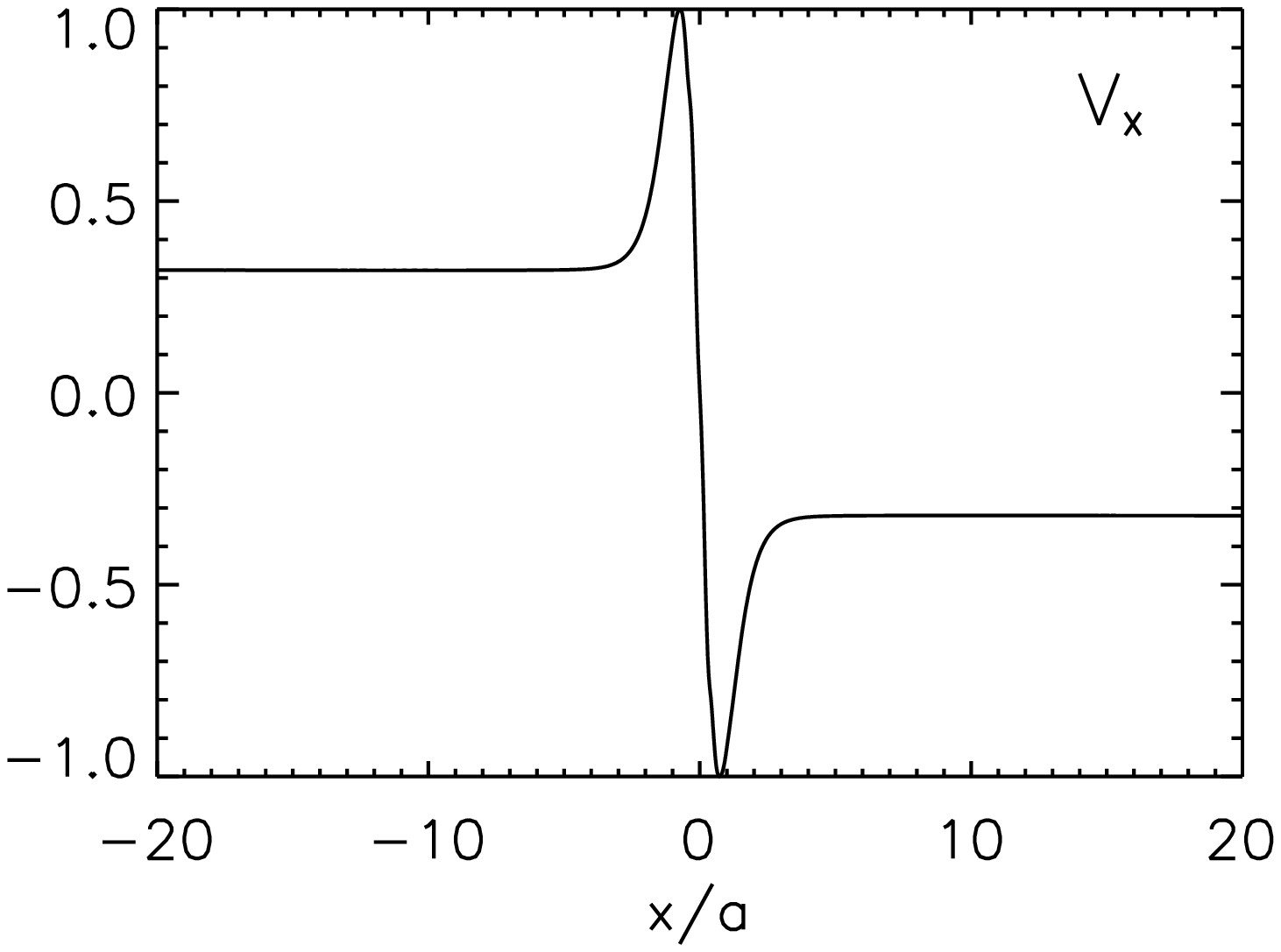} 
\hspace{-14mm} \includegraphics[width=72mm]{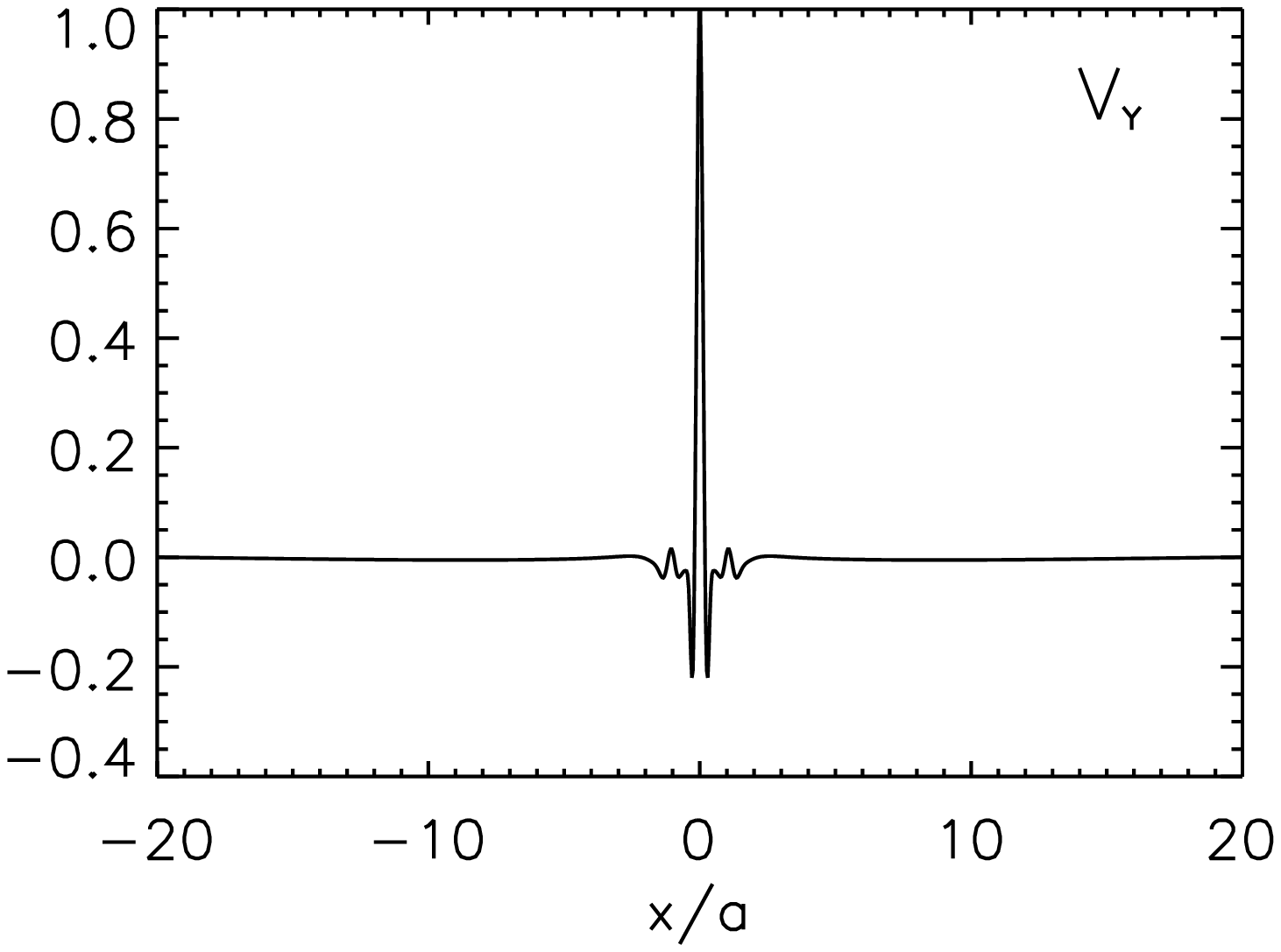} 
  }
\caption{
Numerical eigenmodes of the linear phase for the standard
run with $\sigma_0=1$, $\beta_0=1$ ($c_A=0.5$) and $\bar{S}=10^6$. Here we plot the fields
$B_x$, $v_x$ and $v_y$ for $y=0$ as a function of $x/a$, normalizing all quantities
against their maximum. The shape is modulated with $\cos(ky)$ or $\sin(ky)$ in the
direction parallel to the current sheet.}
\label{fig:eigenmodes}
\end{figure*}


Once more, this classical MHD analysis is expected to apply unchanged in the relativistic case. If a value of the macroscopic Lundquist number $S$ is assigned, and correspondingly $\eta = Lc_A/S$ is redefined according to the value of $c_A$, the maximum ideal tearing instability rate as measured in terms of the light crossing time $\tau_c=L/c\equiv L$ is
\be
\gamma_\mathrm{max} \tau_c \simeq 0.6\, c_A.
\label{eq:rates4}
\ee
This corresponds to the wavenumber (normalized against $a$)
\be
k_\mathrm{max} a \simeq 1.4 \, S^{-1/6},
\label{eq:rates5}
\ee
thus the peak in the dispersion relation $\gamma (ka)$ is expected to decrease for increasing values of $S$ \citep[see the figures in][]{Pucci:2014}.

This is a novel result potentially applicable to any strongly magnetized plasma (provided the fluid description is applicable, of course): contrary to expectations due to previous studies \citep{Komissarov:2007b}, where only the standard case was extended to relativistic plasmas as discussed in section~\ref{sect:standard}, the relativistic tearing mode may indeed occur on light crossing timescales. This result applies provided we assume extremely thin current sheets, with $a/L\sim S^{-1/3}$, or thinner, as initial equilibrium configurations. In particular, for very strong magnetized plasmas in which $c_A\simeq 1$, we expect an instability rate as fast as $\gamma_\mathrm{max}\sim \tau_c^{-1}$, as needed to explain the impulsive magnetic energy conversions observed in many astrophysical sources.

\section{Numerical results: linear phase}
\label{sect:results_linear}

Let us now discuss simulations of the linear phase of the relativistic tearing instability, aimed at confirming numerically the analytical predictions found in the previous section. First we verify the expectations for the standard instability, described in section~\ref{sect:standard}, by checking the dependence of the exponential growth rate with the Lundquist number; then we proceed to investigate numerically the relativistic ideal tearing mode. In both types of simulations we choose to adopt the standard values 
\be
\sigma_0=1, \, \beta_0=1 \, \Rightarrow \, c_A=0.5,
\ee
the value of $c_A$ being the most important parameter, the one expected to directly affect the growth rate $\gamma$.

The compressible, resistive relativistic MHD equations in conservative form, as in section~\ref{sect:model}, are solved with the ECHO code in a rectangular ($x-y$) domain 
\be
[-20a,+20a] \times [0, L_y],
\ee
using $1024\times 512$ computational cells (a higher resolution along $x$ is required due to the sharp gradients for $|x|<a$). Zeroth-order extrapolation is applied at $x=\pm 20a$, while periodic boundary conditions are set along the $y$ direction, parallel to the sheet. Invariance along $z$ will be assumed, thus $\partial/\partial z \equiv 0$ in the MHD equations, though $z$ components of vector fields are retained (we recall the initial equilibrium itself has $B_z\neq 0$), so our simulations are of 2.5-D type.

\subsection{The \emph{standard} tearing instability case}

In order to study the linear phase, the sheet length is chosen to contain a single mode, thus $L_y = 2\pi/k$. Here we normalize all lengths against $a=1$. For a given value of the Lundquist number $\bar{S} = ac_A/\eta$, we select the fastest growing mode wavenumber $k$ approximately as
\be
\bar{k} \equiv ka = \bar{S}^{-1/4}, \quad \gamma\bar{\tau}_c \simeq 0.3\, \bar{S}^{-1/2},
\ee
where $\gamma$ is the corresponding expected growth rate predicted by the theory in section~\ref{sect:standard} for the chosen value of $c_A$.

Initial perturbations are applied only to the magnetic field in the plane $x-y$, whereas velocity and electric field are set to zero. We choose a perturbation of sinusoidal shape in $y$, with the assigned wavenumber $k$, and concentrated around the current sheet using hyperbolic functions of argument $x/a$. A suitable divergence-free form with the desired properties is
\begin{align}
\delta B_x & = \varepsilon B_0 \, \cos (\bar{k} \, y/a)\, \mathrm{sech} (x/a),  \nonumber \\
\delta B_y & = \varepsilon B_0 \,  \bar{k}^{-1} \, \sin (\bar{k} \, y/a)\, \tanh (x/a)\, \mathrm{sech}(x/a),
\label{eq:b1}
\end{align}
where $\varepsilon=10^{-4}$ sets the amplitude of perturbations. Five runs are performed, for $\bar{S}=10^4$, $\bar{S}=3\times 10^4$, $\bar{S}=10^5$, $\bar{S}=3\times 10^5$, and $\bar{S}=10^6$.

In each run, after a rapid transient stage, the eigenmodes characteristic of the tearing mode instability set in and perturbations start to grow exponentially, with rates approximately as those predicted by the theory. Fig.~\ref{fig:eigenmodes} shows the profiles along $x$ and for $y=0$ of some of the perturbations ($B_x$, $v_x$, $v_y$), normalized to their corresponding maximum, at a time during the linear phase of the instability. Notice the similarity with the usual eigenmodes for the classical case, as reported in \cite{Landi:2015}. Here we also observe a smooth inflow from the boundaries at $x=\pm a$, as expected for a SP-type steady reconnection scenario. The shape of these eigenmodes (we recall that a sinusoidal modulation applies along $y$) is roughly the same for all values of the Lundquist number and does not change when the thin current sheet case $a/L\sim S^{-1/3}$ is considered.

\begin{figure}
\centerline{
\hspace{-5mm}
\includegraphics[width=100mm]{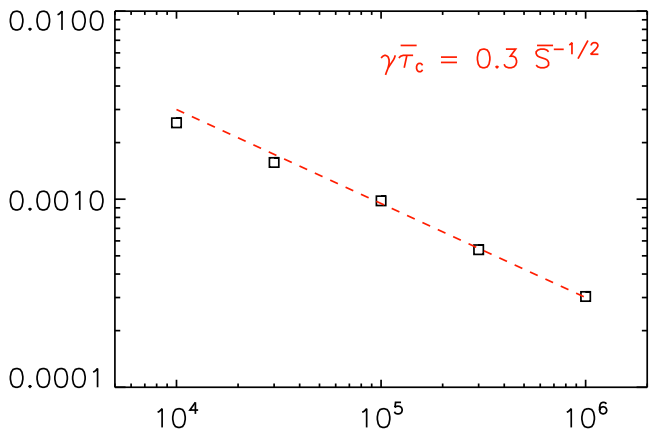}
\vspace{-6mm}
 }
\caption{
Verification of the law $\gamma\bar{\tau}_A\simeq 0.6\,\bar{S}^{-1/2}$, here expressed as $\gamma\bar{\tau}_c\simeq 0.6 c_A\,\bar{S}^{-1/2}$, for the relativistic tearing instability in the standard case with $c_A=0.5$. The red dashed line is the expected behavior. Errors on estimates of $\gamma$ from simulations are approximately of the few percent, within the box symbols.
}
\label{fig:standard}
\end{figure}

The canonical $\bar{S}^{-1/2}$ behavior of the growth rates of the relativistic tearing instability is verified in Fig.~\ref{fig:standard}, where the boxes represent the estimates for the $\gamma\bar{\tau}_c$ values corresponding to the linear phase of the simulations, measured as the growth in time of the maxima of the $B_x\equiv \delta B_x$ perturbations over the whole domain. The accuracy in such estimates is of a few percent, say $<5\%$ (see also Fig.~\ref{fig:growth}), and we notice that the expected rates along the red dashed line are all within $\sim 10\%$ of the measured values. For $\bar{S}=10^6$ we doubled resolution along $x$, otherwise numerical resistivity overcomes the physical one, leading to an effective value of the Lundquist number smaller than expected and to the saturation of the curve. 

We thus conclude that the relativistic MHD theory of the tearing instability proposed in section~\ref{sect:standard} is correct, the only change with respect to the non-relativistic case being the increased inertia in the definition of the Alfv\'en speed. Instability growth rates, measured on the light crossing time $\bar{\tau}_c$ of the current sheet half width $a$, decrease with $\bar{S}$ as in the standard classic case, hence, approaching ideal MHD in the limit of very large $\bar{S}$, as pertaining to many astrophysical scenarios, such rates remain unsatisfactory small.


\subsection{The ideal tearing instability case}

As discussed in section~\ref{sect:ideal}, the decrease of the instability growth rates with the Lundquist number may be avoided by renormalizing all quantities against a macroscopic length $L\gg a$. In particular, by choosing
\be
a=S^{-1/3}L,
\ee
the maximum rate is $\gamma\tau_A \simeq 0.6 \Rightarrow \gamma\tau_c \simeq 0.6 c_A$, of the same order of the inverse of the macroscopic light crossing time $\tau_c=L\gg \bar{\tau}_c$. In the following we set $S=Lc_A/\eta=10^6$, a value high enough to reach approximately the asymptotic regime. For this value we have $a=0.01L$, and the fastest growing mode is expected at the wavenumber
\be
\bar{k}\equiv ka \simeq 1.4 \, S^{-1/6} =0.14.
\ee
This and similar values have been tested numerically in our single-mode analysis. Notice that even if $S$ is quite large, we do not need to increase the numerical resolution as the grid spacing $\Delta x \propto a$ is much smaller than in the previous case, and numerical dissipation is expected to be a problem only for much higher values of $S$.


\begin{figure}
\centerline{
\hspace{-5mm}
 \includegraphics[width=100mm]{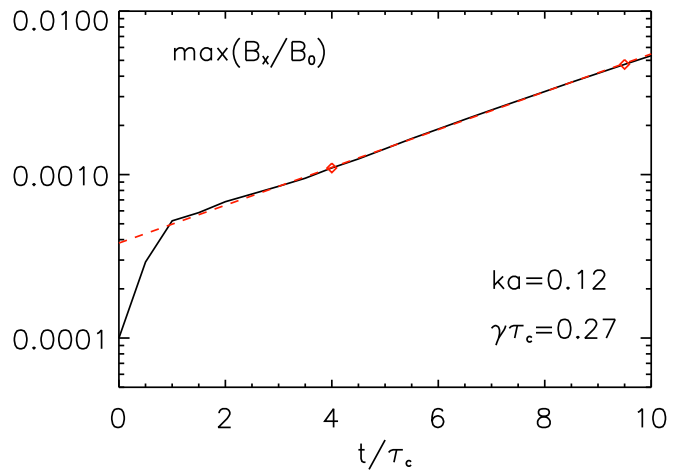}
 }
\caption{
Growth of the first mode of the relativistic ideal tearing instability, for the standard
run with $\sigma_0=1$, $\beta_0=1$ ($c_A=0.5$), and $S=10^6$. The amplitude of
the Fourier transform of $B_x/B_0$ along $y$ (averaged in $x$), corresponding to $\bar{k}\equiv ka=0.12$ 
(the fastest growing mode), are plotted against time in light units $\tau_c=L$.
The red dashed line correspond to the best fit with an exponential growth with $\gamma\tau_c=0.27$,
fit performed within the range indicated by the red symbols. For $t>10\tau_c$ other modes grow
out of the main one and the nonlinear phase begins.
}
\label{fig:growth}
\end{figure}



\begin{figure}
\centerline{
\hspace{-5mm}
 \includegraphics[width=95mm]{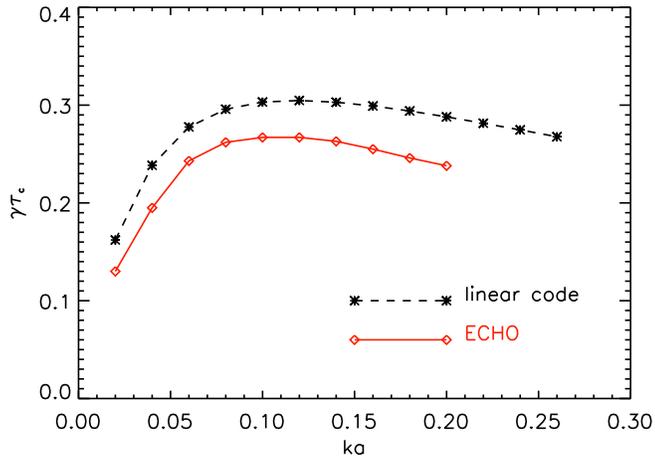}
  }
\caption{
Dispersion relation for the standard run with $\sigma_0=1$, $\beta_0=1$ ($c_A=0.5$), and $S=10^6$,
for  ideal tearing settings with $a=S^{-1/3}L=0.01L$. The red solid line corresponds
to the results obtained with the ECHO code for relativistic MHD, whereas the black dashed line
is computed by using a high-order, linearized code for classical MHD. In that run we use four times the standard
mass density to achieve $c_A=0.5$, and the results exactly match the analytical curve for the ideal tearing instability.
}
\label{fig:dispersion}
\end{figure}


Fig.~\ref{fig:growth} shows the growth of the $B_x$ perturbation as a function of time in the case of the most unstable mode. We find that this does not occur precisely at $\bar{k}=0.14$, but rather for $\bar{k}=0.12$, since we are not sufficiently within the asymptotic regime of very large values of $S$. After an initial stage of rearrangement of the growing perturbations, the plot in Fig.~\ref{fig:growth} clearly shows an exponential growth as $\sim\exp(\gamma t)$, with a fit for $\gamma\tau_c=0.27$ (red dashed line) between the values indicated with asterisks. Errors in the estimate are basically just due to the choice of the fitting points and we can safely assume an accuracy of $2-3\%$ in the measure of $\gamma$, for times small enough to remain in the single-mode linear regime, where mode coupling can be still neglected. 

By selecting different values of $\bar{k}$ around $\bar{k}=0.12$, and by adjusting the domain accordingly, we computed the curve $\gamma(\bar{k})$ numerically and compared it with the theoretical dispersion relation for the relativistic ideal tearing instability. In Fig.~\ref{fig:dispersion} we plot the dispersion relation computed with ECHO (red solid line) with the one obtained by using the linearized version of the compact-Fourier high-order MHD code employed in \cite{Landi:2015} (black dashed line). For consistency, the classical MHD run refers to $c_A=0.5$, obtained by using $B_0=1$ and $\rho_0=4$, and matches exactly the theoretical curve from the linear analysis. 

We believe that the discrepancies (about $10\%$ around the peak for $\bar{k}=0.10-0.12$) observed here between ECHO and the linear code are not due to the physical approximations in our theoretical analysis of the relativistic tearing mode, but rather to the diffusion of the equilibrium magnetic field during evolution. This is enforced by the fact that even in the fully nonlinear MHD case of our previous paper, when diffusion of $\bm{B}_0$ was not removed from the induction equation, growth rates were found to be smaller by the same amount as in the present study. For further details on the classical MHD code and deeper technical discussions see \cite{Landi:2005,Landi:2008}.

Other values of $S$, up to $10^8$, have also been tested, confirming that the maximum growth rates always remain around $\gamma\tau_c \simeq 0.3$, independently on $S$, thus we can safely conclude that we have reached the asymptotic regime of the ideal tearing instability \citep{Pucci:2014}. On the other hand, the peak of the dispersion relation shifts to smaller values of $ka$, approximately as the expected relation of equation~(\ref{eq:rates5}) (for $S=10^8$ we find $\bar{k}_\mathrm{max}\simeq 0.07-0.08$,  to be compared to the expected value of $0.065$). However, we iterate here that such high values of $S$ require an adequate number of grid points to resolve the resistive layer, which is always smaller than the current sheet width (their ratio decreases with $S$ as $S^{-1/2}/S^{-1/3}=S^{-1/6}$), so we prefer not to investigate cases with $S>10^6$ in any further detail.

The theoretical dispersion relation predicted from our analysis of the relativistic tearing instability in thin current sheets with $a=S^{-1/3}L$ is thus confirmed. For higher magnetization and lower plasma beta parameters the Alfv\'en speed of the background plasma, defined in equation~(\ref{eq:ca}), approaches the speed of light, so that the tearing instability occurs on timescales which are comparable to the ideal ones, namely the light crossing time corresponding to the macroscopic scale $L$. Additional simulations with different values of $\sigma_0$ and $\beta_0$ have been performed, and the theoretical prediction $\gamma\tau_A\simeq 0.6 =\mathrm{const}$ for the linear growth of the ideal tearing mode, \emph{independently} of the single plasma parameters, is fully confirmed (see next section).

\section{Numerical results: nonlinear phase}
\label{sect:results_nonlinear}


\begin{figure*}
\centerline{
  \hspace{4mm}   \includegraphics[width=68mm]{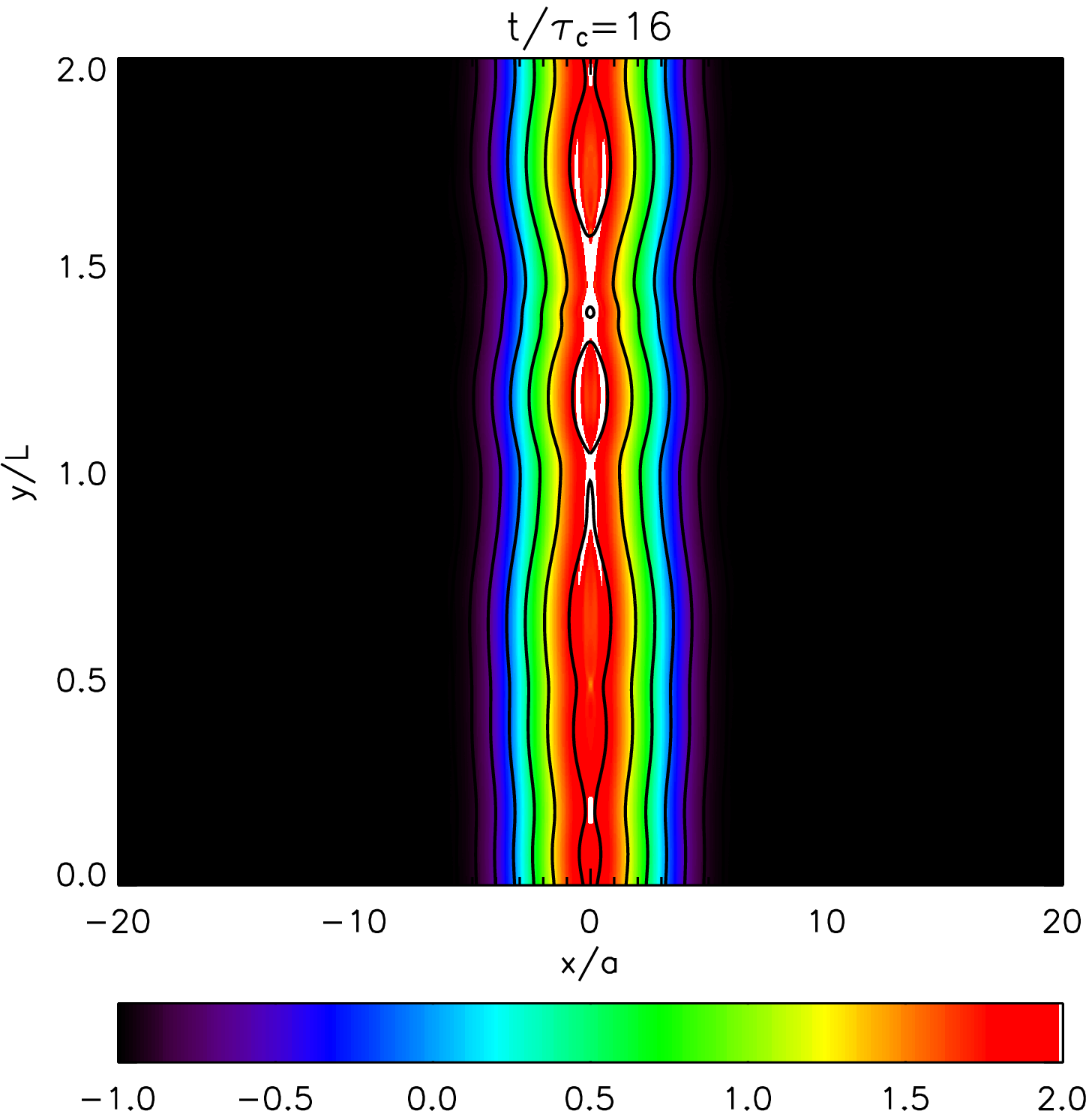}
\hspace{-11mm} \includegraphics[width=68mm]{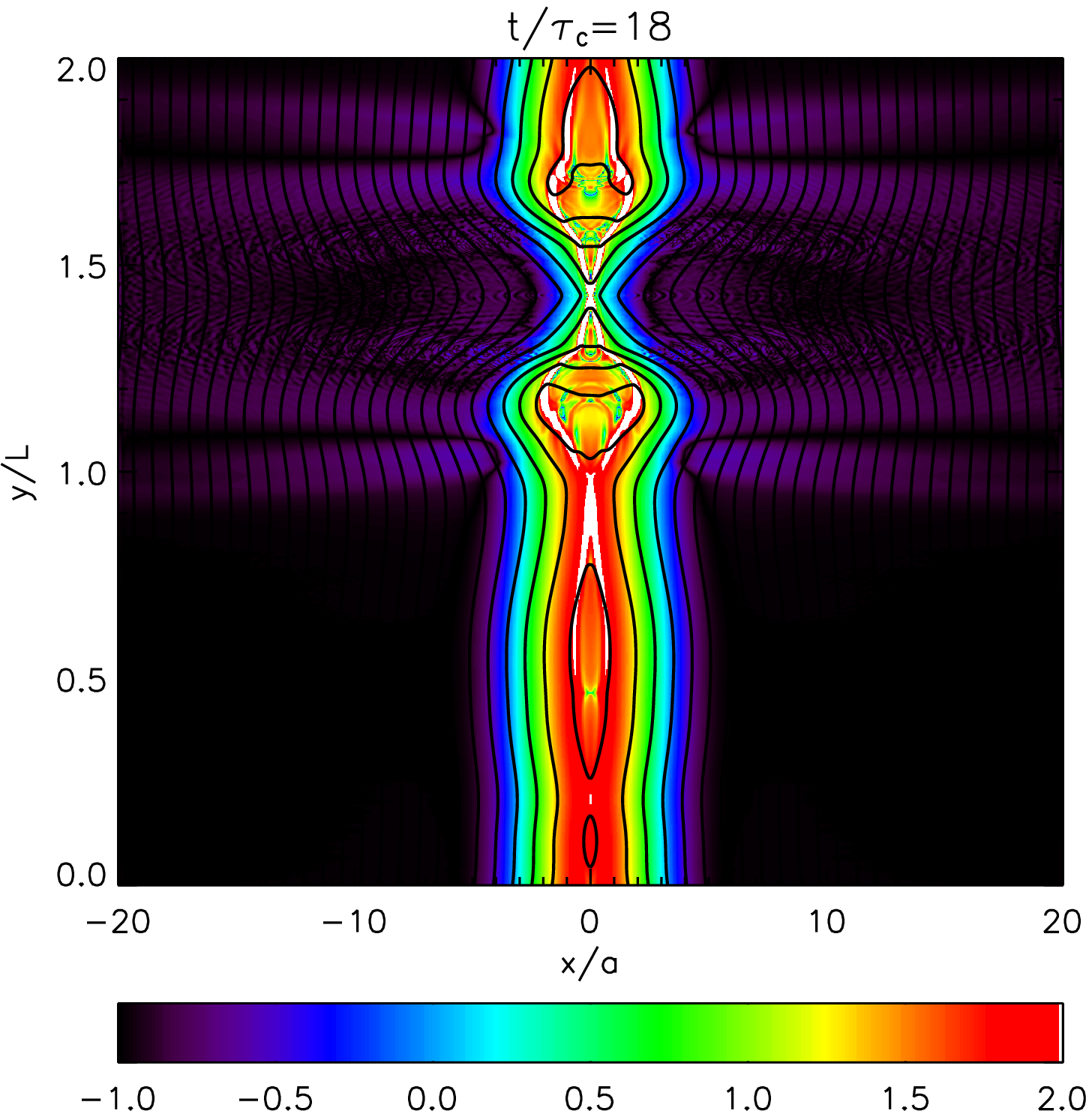}
\hspace{-11mm} \includegraphics[width=68mm]{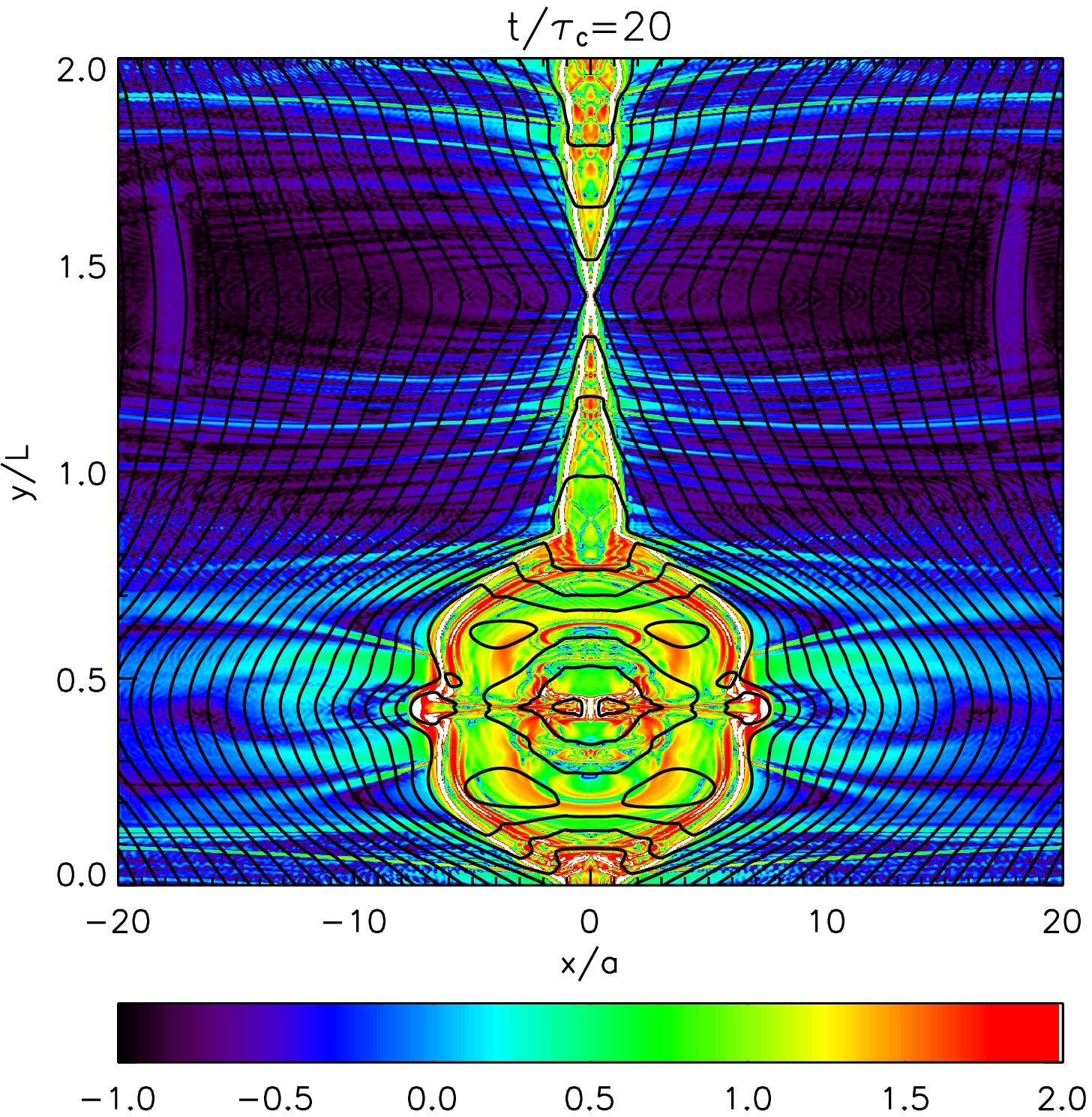}
  }
\caption{
Images of fieldlines and $|(\nabla\times\bm{B})_z|$ in logarithmic scale (normalized to $B_0/L$ and in color scale) at time $t=16\tau_c$, $t=18\tau_c$, and $t=20\tau_c$ in the nonlinear phase of the instability. We employ standard parameters $\sigma_0=1$, $\beta_0=1$ ($c_A=0.5$), and $S=10^6$, for  ideal tearing settings with $a=S^{-1/3}L$. We recall that for the assumed Lundquist number the current sheet is very thin, as $a=0.01L$, so images are clearly not scaled uniformly in $x$ and $y$.
}
\label{fig:JAz}
\end{figure*}


Let us now investigate the fully nonlinear regime of the relativistic ideal tearing instability. We analyze the results of a simulation in which the usual initial force-free configuration in equation~(\ref{eq:b0}) is perturbed by $m_\mathrm{max}=10$ modes, each of them defined as in equation~(\ref{eq:b1}) with a different wavenumber $k_m$ for $m=1,\ldots,m_\mathrm{max}$ and a random phase for each $k$, assuming a root mean square value of magnetic field perturbations of $\epsilon = 10^{-4}$. As in the previous section we use $c_A=0.5$ ($\sigma_0=\beta_0=1$), $S=10^6$, $a=S^{-1/3}L=0.01L$. Here we choose $L_y=2L$, so that along the current sheet we excite modes from $kL=\pi$ to $kL=10\pi$, with a maximum growth rate expected for the mode $k_\mathrm{max}\, a \simeq 0.12$ or $k_\mathrm{max} \,L\simeq 12$, approximately near the center of the selected range in $k$, to allow mode-coupling and both direct and inverse cascade. For this run we employ 1024 grid points in both directions.

As the simulation proceeds, we monitor the growth of each Fourier modes as in \cite{Landi:2015}. We find that the fastest growing modes are those with $m=3$ and $m=4$ (from $m= L_y / \lambda_\mathrm{max}$), in perfect agreement with theory. However, mode coupling soon sets in and we observe the growth of both shorter and longer wavelength modes. In particular, the plasmoids (magnetic islands) arising from the tearing instability grow nonlinearly and merge, and we end up with a state dominated by a single, large plasmoid, and by one well-defined X-point where a secondary reconnection event is about to start. Since we employ periodic boundaries in the $y$ direction, the other phenomenon expected to occur in the nonlinear stage of fast magnetic reconnection, namely the super-Alfv\'enic expulsion of plasmoids from the reconnection sites along the current sheet \citep{Loureiro:2007}, cannot be observed in our simulations.

Perturbations of all quantities are observed to grow in time, exponentially during the linear phase as shown in Fig.~\ref{fig:growth} for $B_x$, then to an even faster rate due to mode coupling, while soon we observe saturation because of fast dissipation at increasingly smaller scales. Given the level of the initial perturbations, the whole process is very rapid, as it completes within the first $\Delta t \simeq 20\tau_c$ macroscopic light crossing times. 


\begin{figure*}
\centerline{
\hspace{4mm}     \includegraphics[width=68mm]{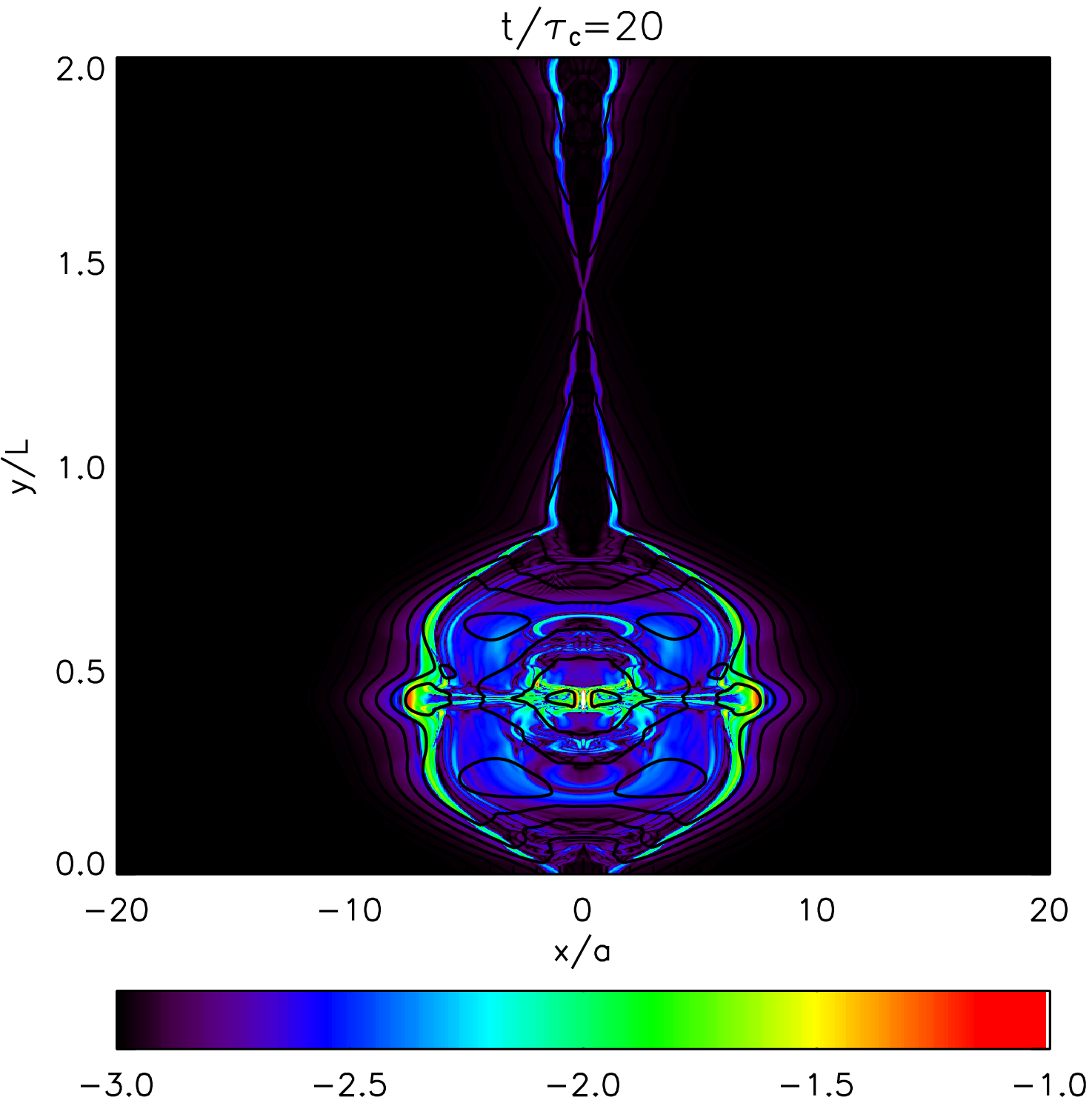}
\hspace{-11mm} \includegraphics[width=68mm]{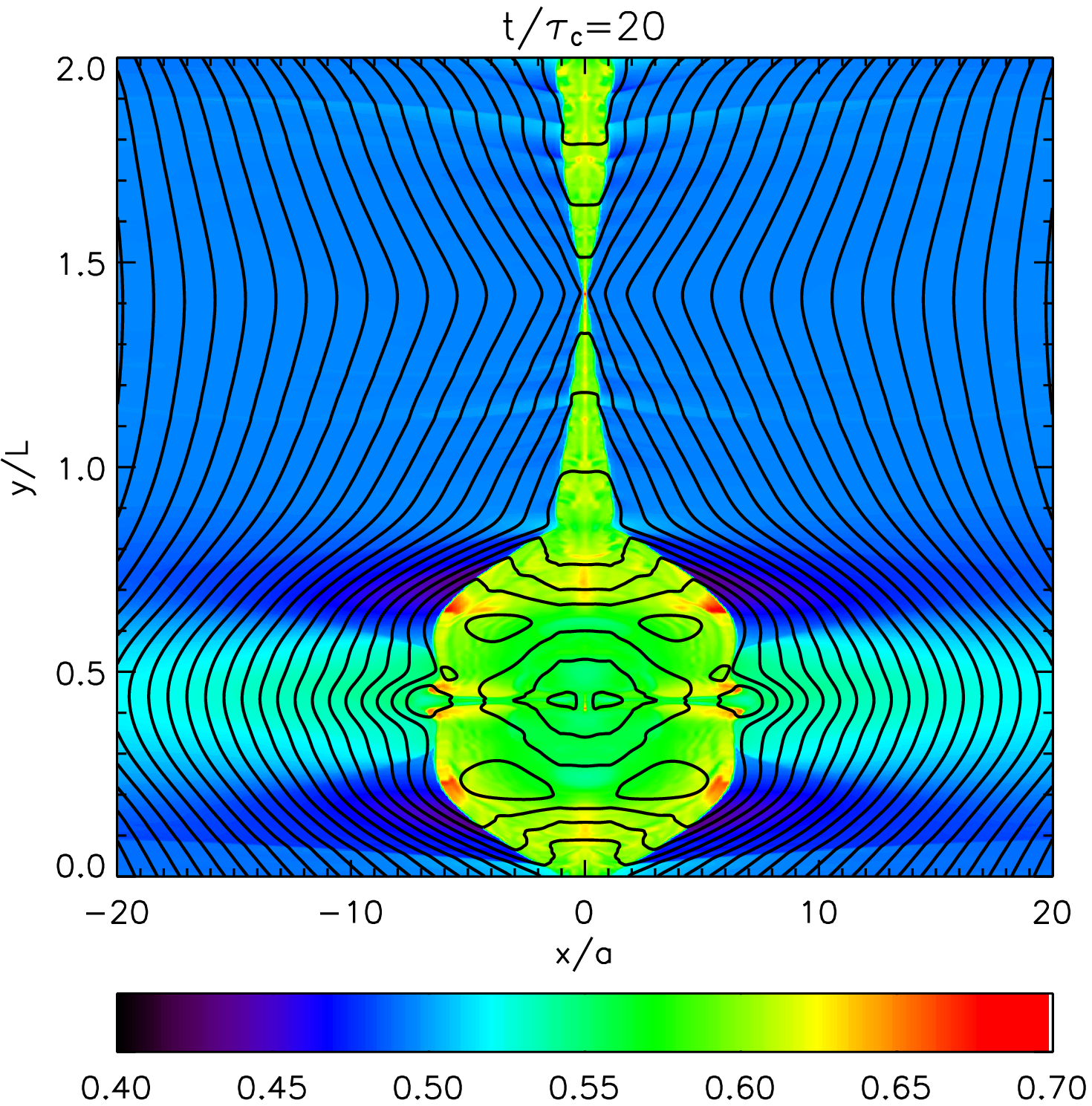}
\hspace{-11mm} \includegraphics[width=68mm]{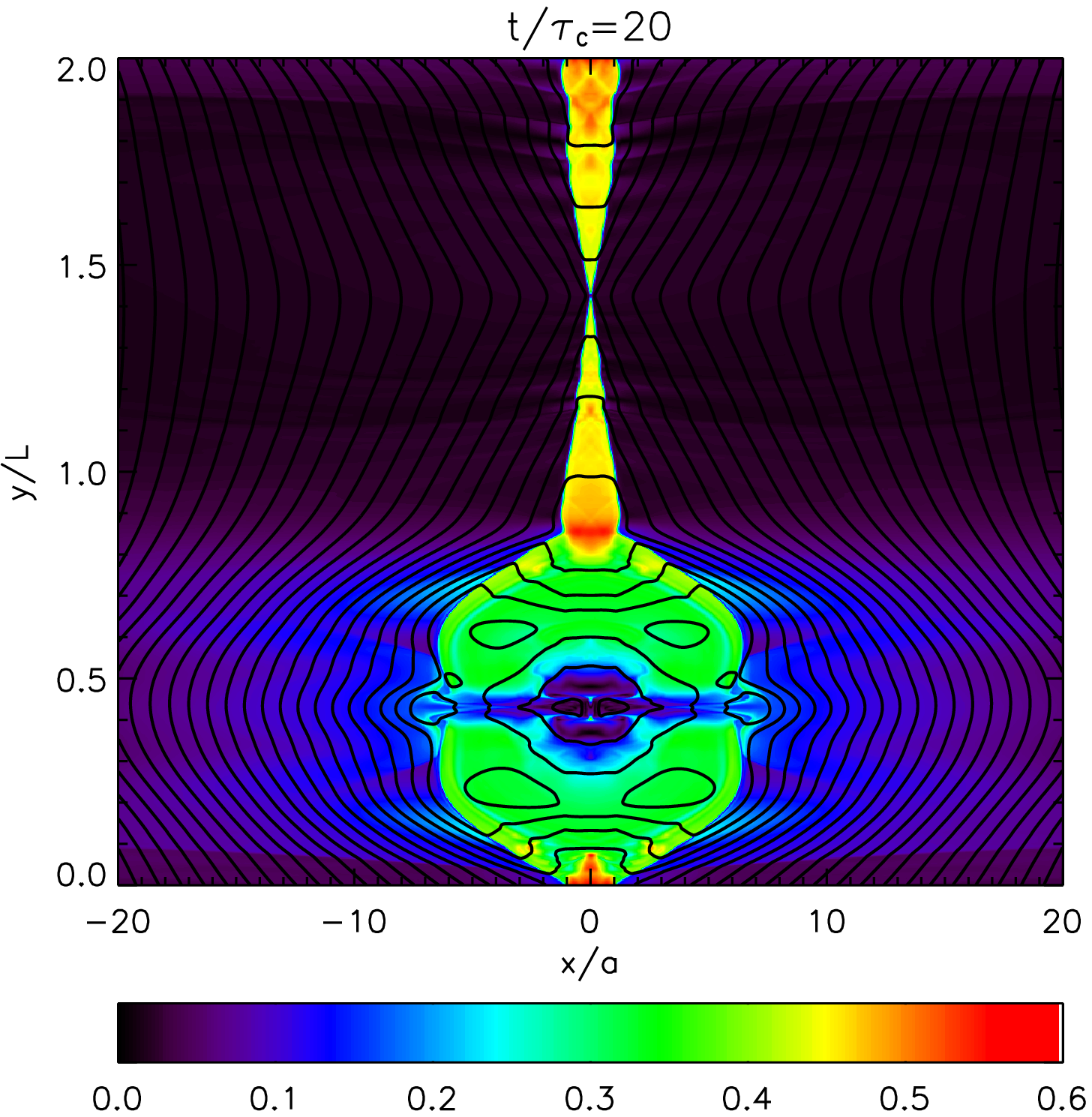}
  }
\caption{
Images of various quantities at $t=20\tau_c$, corresponding to the situation in the right panel of Fig.~\ref{fig:JAz}. From left to right: the quantity $|\bm{E}\cdot\bm{B}|$ in logarithmic scale (saturated for $10^{-3}$ and $10^{-1}$, the maximum is $\simeq 0.25$); the plasma temperature $p/\rho$; the magnitude of the velocity $|\bm{v}|$. We recall that $S=10^6$, hence $a=0.01L$.
}
\label{fig:t20}
\end{figure*}


The distortion of magnetic fieldlines and the intensity of the electric current density $|(\nabla\times\bm{B})_z|$ (here we prefer to use the classical definition, directly related to the topology of the magnetic field, see equation~(\ref{eq:ohm}) for the complete Ohm equation) are shown in Fig.~\ref{fig:JAz} at nonlinear times $t=16\tau_c$, $t=18\tau_c$, and $t=20\tau_c$. Lower and upper cuts are set for the current, here shown in logarithmic scale (we saturate at $10^{-1}$ and $10^2$ in $B_0/L$ units), so that black regions contain even smaller values where the current is negligible (down to $\simeq 10^{-6}$), while isolated white sheets and spikes contain currents with values up to a few hundreds of $B_0/L$. 

We notice the inverse cascade leading to a $m=1$ dominant mode, with a clearly defined, macroscopic X-point at $y\simeq 1.5L$, and a large plasmoid  located between $y=0$ and $y=0.8$, where the magnetic field is structured in a complex pattern of sub-islands and smaller current sheets, due to the recurrent merging events. These two major features are separated by $\Delta y = L_y/2=L$ along the current sheet, whereas their individual locations are casually determined by the random initial conditions.

Regions of high current density also corresponds to narrow structures where $\bm{E}\cdot\bm{B}\neq 0$, thus where particles could be accelerated by non-ideal electric fields along magnetic fieldlines. This is apparent in Fig.~\ref{fig:t20}, where that quantity is shown in logarithmic scale (left panel). Notice that such processes are likely to be more efficient along the boundaries of the current sheet and inside the large plasmoid. In the central panel we show the temperature $T\propto p/\rho$ (here in units of the Boltzmann constant divided by the average particle mass), and we can see that the plasma is heated everywhere in the current sheet, though not dramatically (say $20\%$ on average with respect to the equilibrium value). Finally, in the right panel we show an image of the magnitude of the bulk flow velocity $\bm{v}$. 

The latter is definitely the most interesting plot, since we clearly recognize a Petschek scenario \citep{Petschek:1964}, with high-speed jets originating from the X-point, channeled into an exhaust limited by slow shocks, and finally feeding the plasmoid, which forms just beyond the site where the flow shocks and eventually slows down. Velocities in jets are just mildly relativistic, in this simulation we measure a maximum value of $\simeq 0.55$, precisely when the jets terminate (basically the transition between red and green in the figure). These results appear to confirm what found in previous resistive relativistic MHD simulations \citep{Watanabe:2006,Zenitani:2010,Zanotti:2011}. We recall that, compared to the cited works, in the present simulation the tearing instability develops and saturates on very fast (ideal) timescales. Moreover, here we employ a realistic high value of the Lundquist number ($S=10^6$), and above all we do not assume any enhanced resistivity within the reconnecting region.


\begin{figure}
\centerline{
\hspace{-10mm}
  \includegraphics[width=100mm]{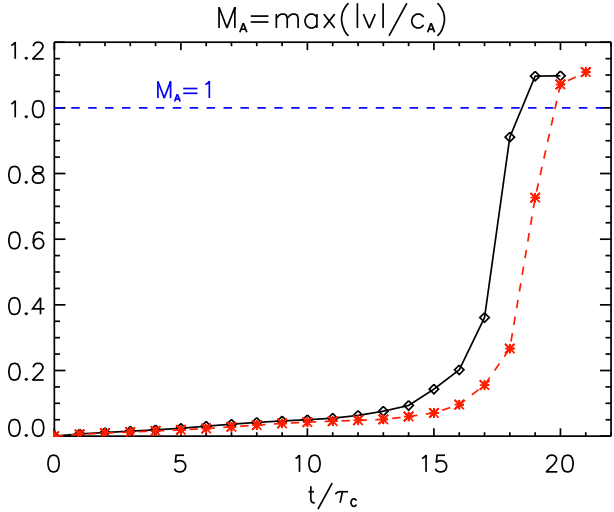}
 }
\caption{
The maximum value over the spatial domain of the Alfv\'enic Mach number $M_A=|\bm{v}|/c_A$ as a function of time, where the Alfv\'en speed $c_A$ refers to the background, unperturbed value of $0.5$. The solid black line (with diamonds) refers to the force-free equilibrium run, the red dashed line (with asterisks) refers to the pressure equilibrium run, the blue dashed line indicates the threshold for super-Alfv\'enic flows.
}
\label{fig:mach}
\end{figure}



\begin{figure}
\centerline{
\hspace{-4mm}
  \includegraphics[width=100mm]{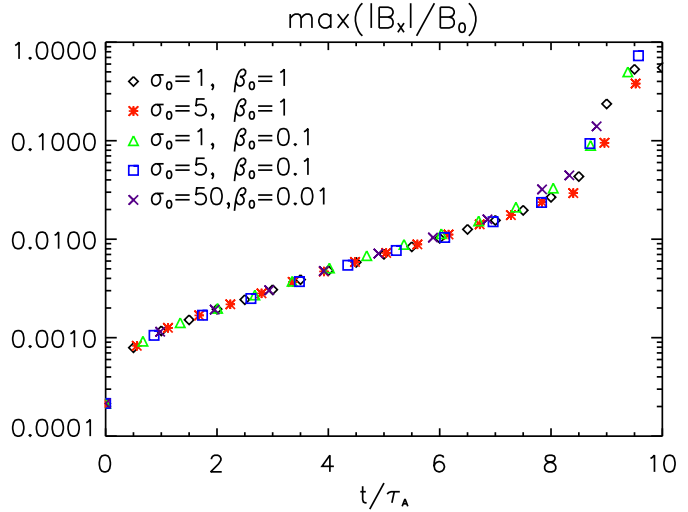}
 }
\caption{
The maximum value over the spatial domain of the $B_x$ component of the magnetic field as a function of time, here normalized to $\tau_A=L/c_A$, for five different simulations. Black diamonds indicate the reference run with $\sigma_0=1, \beta_0=1$ ($c_A=0.5$), red asterisks refer to the case $\sigma_0=5, \beta_0=1$ ($c_A=0.56$), green triangles refer to the case $\sigma_0=1, \beta_0=0.1$ ($c_A=0.67$), blue squares refer to the case $\sigma_0=5, \beta_0=0.1$ ($c_A=0.85$), violet crosses refer to the extreme case $\sigma_0=50, \beta_0=0.01$ ($c_A=0.98$).
}
\label{fig:5runs}
\end{figure}


The overall evolution of the Alfv\'enic Mach number $M_A=|\bm{v}|/c_A$, where the Alfv\'en speed $c_A$ refers to the background, unperturbed value of $0.5$, is shown in Fig.~\ref{fig:mach}, in which $M_A$ is plotted against time. Clearly, a steady growth during the linear phase of the instability is followed by a sudden increase when the nonlinear phase sets in. This corresponds to the stage when jets start to emerge from the major X-point. Though the flow never becomes highly relativistic, as anticipated, it becomes super-Alfv\'enic at $t\simeq 18\tau_c$ (above the blue dashed line in the plot) and the growth saturates after that time. The location with the highest velocity is the shock transition from the exhaust channel to the large plasmoid, as shown in Fig.~\ref{fig:t20} and further discussed below.

In order to verify that this nonlinear configuration is the natural outcome of our ideal tearing runs, we have followed the instability evolution also for an initial equilibrium with $B_z=0$ in equation~(\ref{eq:b0}) and a pressure support determined by
\be
p(x)+\tfrac{1}{2}[B_y(x)]^2=\mathrm{const},
\ee
that is, setting $\zeta=0$ in the formula of \cite{Landi:2015}, while retaining the same values for all other parameters. As expected, since different equilibria do not affect the linear phase of the tearing mode, the nonlinear results are extremely similar: we only notice a slight delay in the final explosive reconnection process and inflation of the large plasmoid, though the maximum velocity peaks around the same level reached in the force-free equilibrium run. The delay is due to the enhanced compressible effects that decrease the linear growth rate of the tearing mode, as already observed in our previous non-relativistic simulations.

With the aim of monitoring the nonlinear evolution at increasing Alfv\'en velocities,  we have run additional simulations by varying the $\sigma_0$ and $\beta_0$ parameters from their reference unit values (for the force-free equilibrium case). In Fig.~\ref{fig:5runs} we plot the maximum value over the spatial domain of the $B_x$ component of the magnetic field, as a function of time and for five different simulations ($c_A=0.50$, $c_A=0.56$, $c_A=0.67$, $c_A=0.85$, and $c_A=0.98$). The black diamonds refer to the standard case ($\sigma_0=\beta_0=1$), whereas violet crosses refer to the last extreme case with $c_A\simeq c$ ($\sigma_0=50, \beta_0=0.01$, see the figure caption for further details). For each simulation, time has been normalized against the corresponding value of $\tau_A=L/c_A=\tau_c (c/c_A)$, so that all curves nicely super-impose almost exactly one over the other. While this is expected in the linear phase, basically by definition since $\gamma\tau_A = \mathrm{const}$ from the theory of the ideal tearing described in the previous sections, it is interesting to notice that the time of the beginning of the nonlinear stage always occurs for $t\simeq 8\tau_A$ (corresponding to $t=16\tau_c$ for the reference run displayed in Fig.~\ref{fig:JAz}), independently of the separate run parameters while depending on $c_A$ alone, and even the subsequent explosive secondary reconnection events seem to follow the same universal path.

Simulations halt when the $B_x$ fluctuations become comparable to the strength of the initial field $B_0$, which means that the fully nonlinear regime has been reached and the largest plasmoid has grown to a considerable size. In all cases we confirm that in the last available output the maximum velocity in jets has become super-Alfv\'enic, against the corresponding value of $c_A$. We deem that the Petschek-like nonlinear configuration described above for the reference case, arising basically unchanged independently on the initial conditions and for different sets of parameters, can be considered as the final stage for this kind of simulations, since the numerical results are no longer reliable for larger times. At that point the dynamics is starting to become completely determined by the chosen boundary conditions: periodicity along $y$ prevents the features to escape and favors the inverse cascade to a quasi-steady single-mode configuration, with a forced, continuous interaction of the jets from the X-point funnels with the large magnetic island. Moreover, across the current sheet, outflow conditions should be applied much farther away than at $x=\pm20a$, due to the vicinity of the plasmoid boundaries, in order to avoid unphysical effects at longer times.

It is interesting to compare the velocity patterns in this Petschek configuration (for the reference run of Figs. ~\ref{fig:JAz} and \ref{fig:t20}, at time $t=20\tau_c$). Assuming an average value of $p/\rho\simeq 0.6$ between the slow shocks and inside the plasmoid, the resulting sound speed in these heated regions is
\be
c_s = \sqrt{\frac{4}{3}\frac{p}{\rho+4 p}}\simeq 0.5,
\ee
basically coincident with the external Alfv\'en speed, whereas we observe a decrease in the local $c_A$, due to the increased inertia and reduced magnetic field. Therefore, the transition of Fig.~\ref{fig:mach} also corresponds to a \emph{fast magnetosonic shock}, precisely that occurring where the exhaust geometry opens up in the larger magnetic island, and the speed is reduced to sub-magnetosonic values. The presence of fast magnetosonic shocks, where jumps of density and velocity occur, is a fundamental ingredient for Fermi-type particle acceleration, obviously lacking in our fluid (MHD), macroscopic description.


\begin{figure}
\centerline{
\hspace{-7mm}
  \includegraphics[width=95mm]{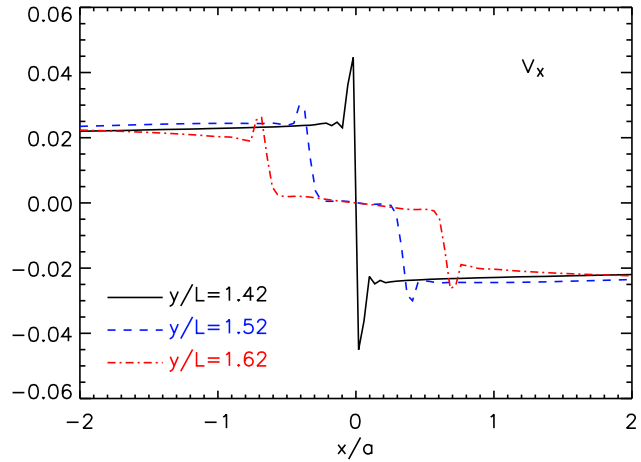}
 }
\caption{
Cuts across the current sheet of $v_x(x,y=\mathrm{const})$ in the exhaust region, from the X-point and beyond, for the indicated values of $y$. The black solid line corresponds precisely to the X-point location where the sign reversal is the sharpest.
}
\label{fig:vcuts}
\end{figure}


As far as the external inflow towards the X-point region is concerned, we measure the $v_x$ component of the velocity as shown in Fig.~\ref{fig:vcuts}, where cuts at constant $y$ are plotted. We notice a rather constant inflow with $|\bm{v}|\simeq |v_x| \simeq 0.025$ from outside (here $v_y\simeq 0$), with an increase to $0.03$ when crossing the slow shocks bounding the exhaust, whereas inside the $v_x$ component approaches zero while $v_y$ becomes dominant and the flow is diverted by $90$ degrees. For $y/L=1.42$ we observe the sharpest transition, precisely at the X-point, with the inflow velocity increasing up to $|\bm{v}|>0.04$ before entering the reconnection region. Excluding the sharp increase just at the X-point, we can conclude that the average Alfv\'enic Mach number of the external inflow is approximately $M_A = |\bm{v}|/c_A \simeq 0.05$, raising to $0.06$ just before entering the exhaust.

The Alfv\'enic Mach number is basically the reconnection rate for a steady Petschek configuration. In that model we have a mild dependence on $S$ as
\be
\mathcal{R} \equiv M_A = \frac{|v_x|}{c_A} = \frac{\pi}{4\ln S}, 
\label{eq:recrate}
\ee
that in our case of $S=10^6$ gives $\mathcal{R} \simeq 0.056$, a value very close to what we measure. This classical result was confirmed for the relativistic case by \cite{Lyubarsky:2005}, who applied it in the limit $c_A\to 1$ as appropriate to magnetically dominated plasmas. Lyubarsky correctly predicted that the outflow velocity would be somehow limited in its growth by the enhanced relativistic temperature in the exhaust (the inertia of the plasma increases), so that the maximum reconnection rate could be at most that in equation~(\ref{eq:recrate}), as fully confirmed by our simulations.

\section{Discussion and conclusions}
\label{sect:concl}

In the present paper we have analyzed, for the first time both analytically and via numerical simulations, the tearing mode instability within relativistic resistive magnetohydrodynamics. Through a semi-relativistic treatment, we have shown here that the instability precisely behaves as in the classical MHD regime, with a timescale for the fastest growing mode proportional to the square of the Lundquist number defined on the current sheet characteristic thickness. The only difference is that normalization of times (or rates) and Lundquist number must be appropriately performed against the relativistic Alfv\'en speed $c_A$ in equation~(\ref{eq:ca}), where the rest mass term $\rho$ in the denominator of the classical expression is replaced by $e+p+B^2/2$, thus it can easily approach the speed of light $c=1$ in magnetically dominated plasmas. Furthermore, we have shown that the numerical eigenmodes are qualitatively as expected by the linear theory. These results extend to the relativistic MHD regime and to higher values of the Lundquist number the work by \cite{Komissarov:2007b}, who investigated the tearing mode instability in the degenerate force-free limit, the approximation in which the fluid inertia and the flow velocities are neglected in the energy-momentum tensor. 

In addition, we have applied to the relativistic case the analysis of the ideal tearing instability for thin current sheets of inverse aspect ratio $a/L\sim S^{-1/3}$ \citep{Pucci:2014}. We find that in this critical regime the maximum growth rate of the instability becomes independent of the Lundquist number $S=L c_A/\eta$, if this is large enough, and in particular
$$
\gamma_\mathrm{max} \simeq 0.6 \, c_A/L,
$$
where $L$ is a macroscopic scale, namely the current sheet length. This result has been here first derived analytically and then accurately confirmed via numerical simulations of the linear phase of the instability, using a high Lundquist number $S=10^6$, basically in the required asymptotic regime. The full dispersion relation $\gamma=\gamma (ka)$ has been also computed via accurate single-mode simulations, showing that the most unstable wavenumber behaves as predicted ($k_\mathrm{max}a \simeq 1.4 \, S^{-1/6}$ or $k_\mathrm{max}L \simeq 1.4 \, S^{1/6}$). 

Finally, fully nonlinear and multi-mode simulations show the evolution of the ideal tearing instability, with exponential growth of the various modes, the subsequent merging of plasmoids and secondary reconnection events as in  \cite{Landi:2015}. In particular, here we find a stage dominated by a single X-point feeding a major plasmoid through fast magnetosonic jets of Petschek-type, shocking directly where the plasmoid has formed. This result is allowed by the assumption of periodic boundary conditions along the sheet direction, with the consequence that the distance between the X-point and the center of the plasmoid is exactly half of the sheet length. If free outflow was assumed, the large plasmoid could exit the boundary and reconnection from the X-point could lead to the elongated plasmoid chains observed by other authors \citep[e.g.][]{Samtaney:2009,Bhattacharjee:2009}. 

Though not induced by the ideal tearing instability, thus subject to much longer evolution times, similar results have been previously obtained in simulations with locally enhanced values of the resistivity \citep{Watanabe:2006,Zenitani:2010,Zanotti:2011}, while we want to stress that in our simulation the resistivity is uniform over the whole domain. Our findings show that the full evolution from the initial linear phase of the tearing instability up to the Petschek scenario only lasts for $t\simeq 10 L/c$ when $c_A\to c$. Moreover, we fully confirm the prediction by \cite{Lyubarsky:2005}, that the outflow speed is not unbound but is basically limited by the external Alfv\'en speed, due to the increased inertia of the relativistically heated plasma inside the jet exhaust \citep[see also][]{Takahashi:2013}. The Petschek fast reconnection rate of $\mathcal{R} \simeq (\ln S)^{-1}$ is retrieved in our nonlinear simulation.

 For this first investigation of the relativistic tearing instability in thin current sheets we have decided to adopt a reference run with unit values for the magnetization and the plasma beta parameters, and we have studied this case in full details. We have then checked that results do not depend on the initial equilibrium model, as both a force-free configuration and a pressure supported sheet lead to very similar evolutions. In addition, we have followed, from the linear to the nonlinear stage, a few cases with higher magnetization $\sigma_0$ (up to $\sigma_0=50$) and lower beta $\beta_0$ (down to $\beta_0=0.01$), for increasing values of the background Alfv\'en speed, from $c_A=0.50$ up to the extreme case of $c_A=0.98\simeq c$. By normalizing the evolution times against the respective $\tau_A=L/c_A$ values, rather than $\tau_c=L/c$, we have derived a sort of \emph{universal curve} for the growth of fluctuations. The ideal tearing model has thus been fully confirmed as far as the linear phase is concerned, since $\gamma\tau_A\simeq 0.6 = \mathrm{const}$ as predicted by the theory, and even the subsequent explosive nonlinear stage appears to depend on $c_A$ alone.

For the future we plan to perform additional simulations in extremely high magnetized plasmas with $c_A\to c$ and to compare results with other authors \citep{Zanotti:2011,Takamoto:2013}. By relaxing the periodicity condition here imposed along the current sheet we would be able to investigate longer nonlinear evolution times, and possibly the statistical behavior of the multiple O-points and X-points which form during the secondary plasmoid instabilities \citep{Uzdensky:2010}. This will be achieved for the first time in the $a/L=S^{-1/3}$ scenario, rather than for the unphysical Sweet-Parker case adopted in the cited works. Moreover, it is our intention to study via relativistic MHD simulations the dynamical thinning process of a current sheet, in order to verify that the ideal tearing sets in as soon as the critical aspect ratio is reached, as already observed in the non-relativistic case \citep{Landi:2015,Tenerani:2015a}.

Our result that the $e$-folding time of the relativistic tearing instability is of order $\tau\sim L/c$ in strongly magnetized plasmas, independently on the Lundquist number, is of the greatest importance for  high-energy astrophysics. Any source in which dynamical processes lead to the formation of thinning current sheets, where the critical threshold $a/L\sim S^{-1/3}$ may be achieved on fast (advection) timescales, is potentially subject to the ideal tearing instability and thus to spontaneous and rapid reconnection processes, with efficient conversion of magnetic energy into fast motions along the sheets and plasma heating. Particle acceleration is obviously not allowed by our MHD computations, though we do observe the formation of narrow regions of non-ideal electric fields with $\bm{E}\cdot\bm{B}\neq 0$ and the presence of fast magnetosonic shocks, in the fully nonlinear phase, both potential candidates to host particle acceleration processes. In the following we briefly discuss the two most direct and promising applications for high-energy astrophysics.

\subsection{Giant flares in magnetar magnetospheres}

An obvious astrophysical application of our model for fast reconnection is to the magnetosphere of magnetars, where we expect fields as high as $B\sim 10^{15}$~G. The possible presence of so-called \emph{twisted-magnetosphere} complex topologies, currently modeled even in general relativity \citep{Pili:2014,Pili:2015,Bucciantini:2015}, is known to be subject to Maxwell stresses that could easily build up, due to motions in the conductive crust, leading to a sheared configuration where progressively thinning current sheets may form on top of the large equatorial loops (if a dipolar configuration is assumed). Fast reconnection of these current sheets, when a critical threshold in the twist is reached, has been invoked to model the sudden release of energy observed as giant gamma-ray flares \citep[e.g.][]{Lyutikov:2003a,Lyutikov:2006a,Komissarov:2007b,Elenbaas:2016}.

Making reference to the last paper, where the known observed soft gamma-ray repeater events are analyzed in detail, the $e$-folding time and the peak time of the light-curves are in the range $\tau_\mathrm{e}=0.1-1$~ms and $\tau_\mathrm{peak}=1-10$~ms, respectively. These numbers are compatible with our model if one assumes $L\simeq 5R_* \simeq 50$~km, as from the cartoon in their paper, so that we predict $\tau_\mathrm{e}=1/\gamma_\mathrm{max}\simeq 0.6c_A/L\simeq 0.2$~ms, with $c_A\simeq c$. Contrary to the modeling in the works cited above, these timescales are \emph{independent} on the Lundquist number $S$ (thus on microphysics and/or on the details of the reconnection process) in the ideal tearing instability scenario, provided that the current sheet thinning can reach the critical threshold $a/L\sim S^{-1/3}$, so that fast reconnection is triggered on ideal timescales.

\subsection{Gamma-ray flares in the Crab nebula}

Sudden and stochastic gamma-ray flares are observed in the Crab nebula \citep{Tavani:2011,FERMI-Collaboration:2011}, which was considered a reference standard candle in that band before this recent discovery. Such short-duration and extremely luminous bursts challenge the standard view of acceleration of pairs at the pulsar wind termination shock via Fermi processes, since the cutoff energy of synchrotron emission shifts during these events up to $\sim 400$~MeV, well beyond the threshold for electrostatic acceleration within ideal MHD (where $E\sim vB < B$). 

Though a low-magnetized pulsar wind is required to reproduce the gross properties of the Crab nebula dynamics and emission in 2.5-D relativistic MHD simulations \citep{Komissarov:2004a,Del-Zanna:2004,Del-Zanna:2006,Volpi:2007,Camus:2009,Olmi:2014,Olmi:2015}, it has recently become clear that efficient dissipation of magnetic energy is possible if fully 3-D simulations are considered \citep{Porth:2014}. Thus, the wind can achieve higher magnetizations before the shock, and relativistic reconnection is currently becoming the most promising mechanism to explain the observations \citep{Lyutikov:2016a}. In particular, current filaments and sheets with substantial magnetization form in the highly unstable postshock nebular flow, so that we can expect a dynamical process leading to the progressive thinning of current sheets, up to the point at which the ideal tearing mode sets in and dissipation of magnetic energy may indeed occur on very fast, advection timescales, as required by the observations.

The rise time $\tau$ of flares in the gamma-ray light-curves is observed to be of a few days, so in order to apply our fast reconnection model a typical length of a reconnecting sheet should be of the order of $L \sim \tau c_A \sim 0.01 - 0.1 R_\mathrm{TS}$, for a magnetized post-shock plasma with $c_A\simeq c$, where $R_\mathrm{TS} \simeq 10^{17}$~cm is the pulsar wind termination shock radius. These numbers seem to coincide with the size of the small-scale bright and time-varying nebular features observed in optical and X-rays near the termination shock, namely the \emph{knot} and the \emph{anvil} \cite[see the discussion in][]{Tavani:2013}, which are thus very strong candidate sites of reconnecting current sheets and hence the sources of the extremely fast and efficient particle acceleration observed during the flares.

\section*{Acknowledgements}

The authors are grateful to M.~Velli, F.~Pucci, E.~Amato, B.~Olmi, and M.~Lyutikov for stimulating discussions, and to the reviewer, M.~Barkov, for his useful suggestions that helped in improving the paper. LDZ acknowledges support from the INFN - TEONGRAV initiative (local PI: LDZ). NB has been supported by a EU FP7 - CIG grant issued to the NSMAG project (PI: NB).

\bibliographystyle{mn2e}

\end{document}